\documentclass[aps,pre,twocolumn,noeprint]{revtex4-1}
\usepackage{graphicx}
\usepackage{amsmath}
\usepackage{tikz}
\PassOptionsToPackage{caption=false}{subfig}
\usepackage{subfig}
\usepackage[bookmarks=false]{hyperref}
\hypersetup{colorlinks=true, citecolor=blue, urlcolor=blue, linkcolor=blue}

\makeatletter
\renewcommand{\maketag@@@}[1]{\hbox{\m@th\normalsize\normalfont#1}}%
\makeatother

\begin{document}

\title{Counting quantum {{}jumps}: a summary and comparison of fixed-time and fluctuating-time statistics in electron transport}
\author{Samuel L. Rudge}
\author{Daniel S. Kosov}

\address{College of Science and Engineering, James Cook University, Townsville, QLD, 4814, Australia }

\begin{abstract}
{{} In quantum transport through nanoscale devices, fluctuations arise from various sources: the discreteness of charge carriers, the statistical non-equilibrium that is required for device operation, and unavoidable quantum uncertainty. As experimental techniques have improved over the last decade, measurements of these fluctuations have become available.} They have been accompanied by a plethora of theoretical literature using many different fluctuation statistics to describe the quantum transport. In this paper, we overview three prominent fluctuation statistics: full counting, waiting time, and first-passage time statistics. We discuss their weaknesses and strengths, and explain connections between them in terms of renewal theory. In particular, we discuss how different information can be encoded in different statistics when the transport is non-renewal, and how this behavior manifests in the measured physical quantities of open quantum systems. All theoretical results are illustrated via a demonstrative transport scenario: a Markovian master equation for a molecular electronic junction with electron-phonon interactions. {{} We demonstrate that to obtain non-renewal behavior, and thus to have temporal correlations between successive electron tunneling events, there must be a strong coupling between tunneling electrons and out-of-equilibrium quantized molecular vibrations.} 
\end{abstract}

\maketitle

\section{Introduction}

Molecular electronic junctions are based on a simple premise; a generic quantum system, such as a molecule or quantum dot, is chemically bonded to two macroscopic metal electrodes \cite{scheer2010molecular,Xiang2016,thoss-evers-review}. Yet, by holding the entire setup out of equilibrium, either with a voltage bias or thermal gradient, one is able to drive particle current through such junctions and thus design novel electronic devices based on the quantum features of the system: such as thermal junctions with unprecedented thermopower \cite{doi:10.1021/nl400579g}, molecular diodes \cite{diode}, single molecule biosensors \cite{sensor}, and THz range rectifiers \cite{17ghz}. As demand for miniaturisation grows and fabrication techniques improve, it is expected that these molecular scale  devices will feature heavily in the new wave of electronics \cite{Xiang2016}. Alternatively, one may use measurable transport properties to probe the dynamics of the quantum system. In this endeavour, molecular electronic devices excel; the juxtaposition of macroscopic electrodes with a quantum system provide observables that are relatively easy to measure but also yield information on fundamental quantum behavior.

In mesoscopic devices, electron tunneling events occur at random time intervals, due in part to thermal fluctuations in the macroscopic electrodes and also to the inherently stochastic nature of quantum mechanics. Any time-dependent observable, for example the current $\hat{I}(t)$, therefore should not be viewed as a constant, but rather a stochastic, dynamical  variable fluctuating around an average \cite{Blanter2000,nazarov-book}. Since such variables are essential to constructing and describing molecular electronic devices, it is imperative to have a rigorous framework able to calculate and analyse these fluctuations. Broadly speaking, fluctuation statistics of electron tunneling events fall into one of two categories discussed below.

\begin{figure} 
{\includegraphics[scale=0.25]{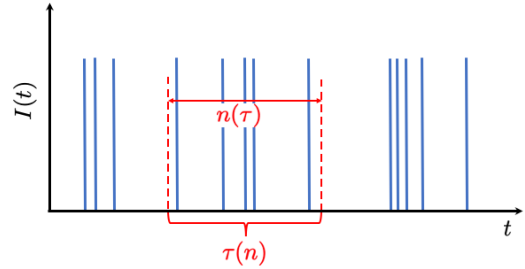}} 
\caption{Schematic of measuring individual electron tunneling events at stochastic points in time. Each electron tunneling event produces one 
``spike" in the electric current measurement. The process could measure the number of electron tunneling events  $n$ over a time interval $\tau$, thus constructing the probability distribution $P(n(\tau))$, or  time $\tau$ it takes for the number of measured quantum jumps to reach value $n$, thus constructing the probability density distribution $P(\tau(n))$. The first probability distribution yields fixed-time statistics  and the second one is connected to fluctuating-time statistics.}
\label{Fig1_schematic}
\end{figure}

Throughout our discussion, we will frequently operate with notions of quantum events and quantum jumps. In the context of quantum transport, the quantum event  is the tunneling of an electron between the molecule and specific electrode, while the quantum jump is the associated transition between two molecular states accompanying the electron tunneling.
Consider an approach in which the number of quantum events $n(t)$ is measured in the time interval $[0,t]$.  Given the stochastic nature of quantum transport, repeat measurements over the same time interval will see $n(t)$ fluctuate about its mean, with probability distribution $P(n(t))$. Alternatively, one could repeatedly measure the time $\tau$ it takes for the number of measured quantum jumps to reach $n$ and construct a probability density distribution $P(\tau(n))$, as we demonstrate in Fig.(\ref{Fig1_schematic}). The first quantity, $n(t)$, is an example of a fixed-time statistic \cite{Levitov1996,Bagrets2003,nazarov-book,Leijnse2008,Kaasbjerg2015,Emary2009,Thielmann2005a,Sukhorukov2000,Thielmann2005,Weymann2008,Carmi2012,Aghassi2008,Braggio2006,Utsumi2006,Park2011,Agarwalla2015,Simine2012}, while $\tau(n)$ is an example of a fluctuating-time statistic \cite{Brandes2008,Saito2016,Rudge2018,Rudge2016a,Rudge2016,PhysRevA.39.1200,Srinivas2010,Thomas2013,Kosov2016,Dasenbrook2015,Albert2011,Albert2012,Kosov2017b,Kosov2018a,Ptaszynski2017,Ptaszynski2017a,Ptaszynski2018}. Considering the important role time-dependent fluctuations have, analysis of quantum fluctuations has therefore been focused on calculating either fixed-time and fluctuating-time statistics, and exploring the relationship between the two \cite{Budini2014,Esposito2010,Ptaszynski2018,Rudge2019}.

We seek in this paper, then, to offer a pedagogical view of quantum fluctuations in single molecule electronic junctions, following the structure outlined by this classification scheme. We aim to write a review of three of the most common statistics: the full counting statistics (FCS) \cite{Bagrets2003,nazarov-book,Leijnse2008,Kaasbjerg2015,Emary2009,Thielmann2005a,Sukhorukov2000,Thielmann2005,Weymann2008,Carmi2012,Aghassi2008,Braggio2006,Utsumi2006,Park2011,Agarwalla2015,Simine2012}, the waiting time distribution (WTD) \cite{Brandes2008,Rudge2018,Rudge2016a,Rudge2016,PhysRevA.39.1200,Srinivas2010,Thomas2013,Kosov2016,Dasenbrook2015,Albert2011,Albert2012,Kosov2017b,Kosov2018a,Ptaszynski2017,Ptaszynski2017a,Rudge2019}, and the first-passage time distribution (FPTD) \cite{Saito2016,Ptaszynski2018,Singh2018,Rudge2019}. This will include a brief history and context for each, an outline of how to calculate the statistic, and a demonstration of the statistic for an example model. Most importantly, since it has been a major focus of quantum fluctuations, we discuss the connections between fixed-time and fluctuating-time statistics as well as the separate and complementary information contained within. This is not a complete review; quantum fluctuations have a broad history in mesoscopic physics and many statistics have been calculated using various theoretical techniques, such as scattering theory, non-equilibrium Green's functions, and Markovian master equations. {{} At all times we restrict our analysis to Markovian master equations, since in this framework a general notation spanning the three different statistics can be developed. Restrictive though this regime is, Markovian master equations are one of a group of techniques in which inelastic scatterings on the molecule can be treated exactly.}

Although, throughout the paper, we illustrate ideas employing examples from quantum transport, all our results should be transferable and accessible to a wider community of theorists working on the dynamics of open quantum systems.

The paper is organised as follows. In Section (\ref{Model Section}) we introduce first the generic framework of Markovian master equations, and also the specific transport scenario we use to demonstrate all statistics: the Holstein model. In Section (\ref{Fixed-time statistics Section}) we introduce a single fixed-time statistic: the FCS. This includes its history, theoretical outline, and some analytic results for the Holstein model. Similarly, Section (\ref{Fluctuating-time statistics Section}) introduces and analyses two fluctuating-time statistics: the WTD and the FPTD. In Section (\ref{Renewal and non-renewal Section}) we explore the connections between them, the conditions under which these connections are true, and what unique practical information we can obtain from each statistic. Section (\ref{Example Section}) contains numeric results for all statistics, calculated from the Holstein model. We conclude with Section (\ref{Conclusions}).

Throughout the paper we use natural units: $\hbar = e = k_{B} = 1$.

\section{Model system} \label{Model Section}

\subsection{General Markovian quantum system}

Time-varying statistical techniques must be applicable to a wide range of quantum transport situations, which we encapsulate in the general scenario of a quantum system weakly coupled to multiple thermal reservoirs. The reduced density matrix of the quantum system $\mathbf{P}(t)$ is assumed to satisfy the master equation:

\begin{align}
\dot{\mathbf{P}}(t) & = \mathbf{L}\mathbf{P}(t). \label{standard ME}
\end{align}

Master equations are an intuitive tool for describing transport through open quantum systems and, as we shall see, provide an easy gateway to fluctuation statistics. {{} The key assumption in Eq.\eqref{standard ME} is that the transport can be approximated as Markovian; bath correlation functions decay quickly and the Liouvillian $\mathbf{L}$ is time-independent or local in time. If this assumption is not satisfied, then the transport is defined by the kernel $\mathcal{K}(t,t')$:

\begin{align}
\dot{\mathbf{P}}(t) & = \int^{t}_{t'}\mathcal{K}(t,t')\mathbf{P}(t'). \label{standard non-Markovian ME}
\end{align}
 
In all our calculations, however, we assume the transport is Markovian and is  governed by Eq.\eqref{standard ME} with time-independent Liouvillian.} In both master equations we have implicitly written the density matrix in superoperator notation $\mathbf{P}(t) = \left[  \text{diagonal elements} , \text{coherences}  \right]^{T}$; from here onwards, though, we will  ignore all coherent superpositions of system states,  which is not a critical approximation for our subsequent {{} mathematical derivations; the derivations can be easily extended to include coherences.} Eq.\eqref{standard ME} then becomes a rate equation written in the orthogonal basis of system states, which has clear physical meaning. $\left\{\mathbf{P}(t)\right\}_{l}$ is the probability that the quantum system occupies state $l$, and may transition from state $l$ to state $k$ via tunneling to or from the thermal reservoirs, with associated rate $\Gamma_{kl}$. The Liouvillian $\mathbf{L}$ is thus constructed according to the system dynamics; the off-diagonals are $[\mathbf{L}]_{kl} = \Gamma_{kl}$ and the diagonals are $[\mathbf{L}]_{kk} = -\sum\limits_{l\neq k}\Gamma_{lk}$.

Eq.\eqref{standard ME} has the general solution 

\begin{align}
\mathbf{P}(t) = e^{\mathbf{L}t}\mathbf{P}(0), \label{General master equation solution}
\end{align}
where $\mathbf{P}(0)$ is the initial system at $t=0$. The master equation Eq.\eqref{standard ME} is assumed to have a unique stationary solution, steady state $\bar{\mathbf{P}}$, which is a null vector satisfying $\mathbf{L}\bar{\mathbf{P}} = 0$. 

As opposed to Eq.\eqref{standard ME}, which we refer to as the $\text{\it{standard}}$ master equation to avoid confusion, the $n$-resolved master equation resolves the transport upon the total number of particles $n$ transferred to the drain:

\begin{align}
\dot{\mathbf{P}}(n,t) & = \sum_{n'}\mathbf{L}(n-n')\mathbf{P}(n,t). \label{n-resolved ME}
\end{align}
Here, $[\mathbf{P}(n,t)]_{k}$ is the probability that the system is in state $l$ at time $t$ and $n$ total particles have been transferred to the drain in the time $[0,t]$. The total number of transferred particles $n$ is sometimes in the literature \cite{Ptaszynski2018} referred to as the jump number, a notation that we adopt here. We consider processes that, at most, add or remove one particle from the drain, so $n-n' = 0,\pm1$. 

Eq.\eqref{n-resolved ME}, consequently, can be written explictly using quantum jump operators that move particles forward to the drain $\mathbf{J}_{F}$ or backwards from the drain $\mathbf{J}_{B}$, alongside an operator containing the remaining dynamics $\mathbf{L}_{0} = \mathbf{L} - \mathbf{J}_{F} - \mathbf{J}_{B}$:

\begin{align}
\dot{\mathbf{P}}(n,t) & = \mathbf{L}_{0}\mathbf{P}(n,t) + \mathbf{J}_{F}\mathbf{P}(n-1,t) + \mathbf{J}_{B}\mathbf{P}(n+1,t). \label{n-resolved ME expanded}
\end{align} 

In Fourier space, Eq.\eqref{n-resolved ME expanded} becomes 

\begin{align}
\dot{\mathbf{P}}(\chi,t) & = \mathbf{L}(\chi)\mathbf{P}(\chi,t), \text{ where }  \label{Chi dependent ME} \\
\mathbf{L}(\chi) & = \mathbf{L}_{0} + \mathbf{J}_{F}e^{i\chi} + \mathbf{J}_{B}e^{-i\chi}. \label{Full Liouvillian Chi}
\end{align}
The infinite set of coupled differential equations has now become a workable problem via the introduction of a counting field $\chi$ \cite{Bagrets2003,Nazarov1999}, where the $\chi$-dependent Liouvillian originates from the Fourier transform of $\mathbf{P}(n,t)$:

\begin{align}
\mathbf{P}(\chi,t) & = \sum_{n}e^{in\chi}\mathbf{P}(n,t), \label{Fourier transformed probability vector} \\
\mathbf{P}(n,t) & = \frac{1}{2\pi}\int_{0}^{2\pi} d\chi e^{-in\chi} \mathbf{P}(\chi,t). \label{Inverse Fourier transformed probability vector}
\end{align}

The initial condition of Eq.\eqref{Chi dependent ME} is $\mathbf{P}(\chi,0) = \bar{\mathbf{P}}$, since at time $t=0$ the open system has already reached the stationary state, and it has the general solution

\begin{align}
\mathbf{P}(\chi,t) & = e^{\mathbf{L}(\chi)t}\bar{\mathbf{P}}. \label{Chi dependent ME solution}
\end{align}

\subsection{Holstein model}

In this section we introduce the specific quantum system used to demonstrate all fluctuation statistics, and describe its dynamics with a Markovian rate equation. 

First, the Hamiltonian of a total system and macroscopic source (S) and drain (D) electrode configuration is 

\begin{equation}
H= H_{M} + H_{S} + H_{D} + H_{T},
\end{equation}
where $H_{M}$ is the molecular Hamiltonian, $H_{S}$ and $H_{D}$ are the source and drain Hamiltonians, and $H_{T}$ describes the molecule-electrode interaction. We consider a molecule described by a single electronic level  interacting with a localised vibration:
\begin{equation}
H_{M}= \varepsilon_0 a^\dag a  + \lambda \omega (b^\dag + b) a^\dag a +  \omega b^\dag b,
\end{equation}
where $\varepsilon_0$ is the molecular orbital energy,  $\omega$ is the vibration frequency, and $\lambda$ is the strength of the electron-vibration coupling. The operator $a^\dagger$  creates an electron in the molecular orbital, while $a$ annihilates an electrom from the molecular orbital. Similarly, $b^{\dagger}\text{ and }b$ serve as bosonic creation and annihilation operators for the phonons. We assume that a strong magnetic field prevents spin-degeneracy in $\varepsilon_{0}$, and so spin is excluded from calculations.

The electrodes are described as a sea of noninteracting electrons:
\begin{align}
H_{S}+H_{D} = \sum_{\alpha = S,D}\sum_{k}\varepsilon^{}_{\alpha,k}a^{\dagger}_{\alpha,k}a^{}_{\alpha,k},
\end{align}
while the molecule-electrode interaction is 
\begin{align}
\label{V}
H_{T} = \sum_{\alpha = S,D}\sum_{k}t^{}_{\alpha,k}(a^{\dagger}_{\alpha,k}a^{}+a^{\dagger}a^{}_{\alpha,k}).
\end{align}
Here, $a^{\dagger}_{\alpha,k}$ and $a^{}_{\alpha,k}$ are the creation and annihilation operators for state $k$ in electrode $\alpha$, and $t_{\alpha,k}$ is the tunneling matrix element. 

{{} The Holstein model is a non-trivial model for transport through single molecules as it incorporates not only the electronic current but also the resulting molecular vibrations, albeit in the simplified harmonic and linear coupling framework. It offers a rich tapestry of physical phenomena, such as unequilibrated phonons, correlations, and dynamical energy flow between electronic and vibrational degrees of freedoms; as such it has already been the subject of many theoretical and experimental papers analysing various aspects of the transport. Experimental measurements on a $\text{C}_{\text{60}}$ molecule, for example, demonstrated a fundamental relationship between electronic hopping and the nanomechanical oscillations of the molecule \cite{Park2000}, which has spawned theoretical work analysing the corresponding FCS \cite{Flindt2005} via the Holstein model. Other important work are early studies of fluctuations in the Holstein model, modelled with quantum master equations. Koch et al., in particular, found that at low voltages a strong electron-phonon coupling suppresses the electronic current \cite{Koch2005}. In this regime the transport is dominated by avalanches of electrons separated by long periods of no transitions, which they found evidence for in the noise \cite{Koch2005} and FCS \cite{Koch2005a}.}

After the Lang-Firsov transformation \cite{Lang1963}, the molecular Hamiltonian is diagonalised to
\begin{align}
H_{M} & = \varepsilon \tilde{a}^{\dagger}\tilde{a} + \omega \tilde{b}^{\dagger}\tilde{b}. \label{LF Hamiltonian}
\end{align}
The new fermionic operators $\tilde{a}^{\dagger}\text{ and }\tilde{a}$, which are obtained from the canonical transformation, create and annihilate electrons in the molecular orbital with shifted energy $\varepsilon = \varepsilon_{0} - \frac{\lambda^{2}}{\omega}$, and similarly for $\tilde{b}^{\dagger}\text{ and }\tilde{b}^{}$. Eq.\eqref{LF Hamiltonian} then has energy eigenvalues $E_{mq} = \varepsilon m + \omega q$ and eigenstates $|mq\rangle$, composed of two quantum numbers: the electronic occupation $m = \left\{0,1\right\}$ and the phonon occupation $q = \left\{0,1,2, \hdots , +\infty\right\}$. 

The probability that the molecule is occupied by $m$ electrons and  $q$ vibrational quanta at time $t$, $P_{mq}(t)$, is described in the Born-Markov approximation by the master equation \cite{Mitra2004}

\begin{widetext}
\begin{eqnarray}
\dot P_{0q}(t) &=& \sum_{\alpha q'} \Gamma^\alpha_{0q,1q'} P_{1q'} (t)  -  \Gamma^\alpha_{1q', 0q} P_{0q}(t),
\label{me1}
\\
\dot P_{1q}(t)&=& \sum_{\alpha q'} \Gamma^\alpha_{1q,0q'} P_{0q'}(t)  -  \Gamma^\alpha_{0q',1q} P_{1q}(t).
\label{me2}
\end{eqnarray}
\end{widetext}

Transitioning between state $|1q\rangle$ to state $|0q'\rangle$ via electrode $\alpha$ occurs with rate 

\begin{equation}
\Gamma^\alpha_{0q',1q} =  \gamma^\alpha |X_{q'q}|^2 \left[1-n_{F}(\varepsilon-\omega (q'-q) - \mu_{\alpha}) \right],
\end{equation}
and likewise for the transition rate between $|0q\rangle$ and $|1q'\rangle$ via electrode $\alpha$:
\begin{equation} 
\Gamma^\alpha_{1q',0q} =    \gamma^\alpha |X_{q'q}|^2  n_{F}\left(\varepsilon+\omega (q'-q)-\mu_\alpha\right).
\end{equation}

The rates depend on the Fermi-Dirac occupation function 
\begin{equation}
n_{F}(E-\mu_{\alpha}) = \frac{1}{1+e^{(E-\mu_\alpha)/T}},
\end{equation}
which in turn is a function of the electrode temperature $T$; the chemical potential of the $\alpha$ electrode $\mu_{\alpha}$; the Franck-Condon factor $X_{qq'}$ 
\begin{align}
X_{qq'} & = \langle q | e^{-\lambda(b^{\dagger}-b)}|q'\rangle;
\end{align}
and the broadening of the electronic level due to electrode coupling $\gamma^{\alpha} = 2\pi|t_{\alpha}|^{2}\rho(\varepsilon)$, where the density of states $\rho(\varepsilon)$ is assumed to be constant.

Although Eq.\eqref{me1} and Eq.\eqref{me2} are an infinite set of coupled differential equations, in practice one may reduce this to finite dimensionality by capping the maximum number of phonons available for transport at some value $N$. The value of $N$ is selected such that $N\omega$ is much larger than all relevant energy scales in the system such as source-drain voltage bias $V_{SD} = \mu_{S} - \mu_{D}$, level broadening $\gamma$, and temperature $T$.  The probability vector $\mathbf{P}(t)$ and the $n$-resolved $\mathbf{P}(n,t)$ therefore both have length $2(N+1)$. In Fourier space, then, the $n$-resolved probability vector is 

\begin{align}
\mathbf{P}(\chi,t) & = \left[\begin{array}{c} P_{00}(\chi,t) \\ P_{10}(\chi,t) \\ P_{01}(\chi,t) \\ P_{11}(\chi,t) \\ . \\ . \\  . \\  P_{0N}(\chi,t) \\ P_{1N}(\chi,t) \end{array} \right],
\end{align}
and the corresponding master equation \cite{Mitra2004} is 
\begin{widetext}
\begin{align}
\dot{P}_{0q}(\chi,t) & = \sum_{q'} \left(\Gamma^{S}_{0q;1q'} + \Gamma^{D}_{0q;1q'}e^{i\chi}\right)P_{1q'}(\chi,t) - \sum_{\alpha q'} \Gamma^{\alpha}_{1q';0q}P_{0q}(\chi,t), \label{P0 rate equation holstein} \\ 
\dot{P}_{1q}(\chi,t) & = \sum_{q'} \left(\Gamma^{S}_{1q;0q'} + \Gamma^{D}_{1q;0q'}e^{-i\chi}\right)P_{0q'}(\chi,t) - \sum_{\alpha q'} \Gamma^{\alpha}_{0q';1q}P_{1q}(\chi,t) \label{P1 rate equation holstein}.
\end{align}
\end{widetext}
The jump operators $\mathbf{J}_{F}$ and $\mathbf{J}_{B}$ are obtained via physical considerations from Eq.\eqref{P0 rate equation holstein} and Eq.\eqref{P1 rate equation holstein}, and are explicitly written in  Appendix  B of our recent paper \cite{Rudge2019}.

When the phonons are forced to relax to equilibrium immediately via an external bath at temperature $T_{V}$, transport is described by a master equation with a much-reduced dimensionality:

\begin{align}
\mathbf{L}_{0} & = \left[\begin{array}{cc} -T_{10} & T_{01}^{S} \\ T^{S}_{10} & -T_{01} \end{array}\right], \label{equilibrated Lo}
\end{align} and jump operators

\begin{align}
\mathbf{J}_{F} & = \left[\begin{array}{cc} 0 & T_{01}^{D} \\ 0 & 0 \end{array}\right], & \mathbf{J}_{B} & = \left[\begin{array}{cc} 0 & 0 \\ T_{10}^{D} & 0 \end{array}\right]. \label{equilibrated jump operators}
\end{align}

The master equation reduces in this way since, for equilibrated phonons, the state probabilities can be separated with the ansatz 
\begin{align}
P_{nq}(\chi,t) & = P_{n}(\chi,t)\frac{e^{-q\omega/T_{V}}}{1-e^{-\omega/T_{V}}} 
\end{align}
and the individual transition rates are 
\begin{align}
T^{\alpha}_{kl} = \sum_{qq'} \Gamma^{\alpha}_{kq;lq'} \frac{e^{-q\omega/T_{V}}}{1-e^{-\omega/T_{V}}},
\end{align}
which sum as $T_{kl} = \sum\limits_{\alpha} T_{kl}^{\alpha}$.

We will use the case of equilibrated phonons as a demonstrative example; since the matrices are all $2\times2$ the fluctuation statistics can all be analytically derived. The stationary state, satisfying $\mathbf{L}\bar{\mathbf{P}} = 0$ and $\bar{\mathbf{P}}_{0} + \bar{\mathbf{P}}_{1} = 1$, is 

\begin{align}
\bar{\mathbf{P}} & = \frac{1}{T_{10}+T_{01}}\left[\begin{array}{c} T_{01} \\ T_{10}\end{array}\right].
\end{align}

This case is also an example of a single-reset open quantum  system: a system where every tunneling to the drain leaves the molecule empty. For single-reset systems $\big(\mathbf{J}_{F} \big)^{k}\mathbf{P} = 0$ and  $\big(\mathbf{J}_{B} \big)^{k}\mathbf{P} = 0$ for any vector $\mathbf{P}$ when $k>1$. 

In comparison, the non-equilibrium case serves as an example of a multiple-reset system; after a tunneling to the drain the molecule may be occupied by zero electrons but any number of phonons. Multiple-reset systems are often difficult to handle analytically due  to larger matrix sizes and system complexity.

\section{Fixed-time statistics} \label{Fixed-time statistics Section}

Early experimental measurements in mesoscopic transport focused on the stationary average electric current $\langle I \rangle$ and, as techniques improved, the zero-frequency noise $S(0)$: essentially the first and second cumulants of the current distribution. Despite the host of information available from these statistics, traditional noise measurement methods are difficult to implement for mesoscopic systems as currents are extremely small. A non-Gaussian current distribution, furthermore, will not be fully described by its average and variance \cite{Bruderer2014,Reulet2003,Bomze2005}, and higher order time-dependent cumulants $\langle\langle I(t)^{k} \rangle\rangle$ must be included: the FCS. Experimentally, obtaining all cumulants of the current distribution requires time-resolved single-electron detection techniques. 

{{} In an early experiment, Reulet et al. \cite{Reulet2003} related measurements of the skewness of voltage fluctuations to the skewness of current fluctuations. Other, more direct, methods for obtaining the higher-order current cumulants have generally used a quantum point contact (QPC) operating as a nearby charge detector. Lu et al. \cite{Lu2003} and Fujisawa et al. \cite {Fujisawa2004}, for example, capacitively coupled a single-electron transistor (SET) to a quantum dot (QD), which at low $V_{SD},T$ can only change occupation by $\pm 1$ electrons. The capacitive coupling ensures that individual tunnelings of electrons changes the charging energy, and thus differential resistance, of the SET by a discrete amount. Armed with a method able to count single-electron transitions, the QPC setup was used to calculate the FCS of a quantum dot \cite{Bomze2005,Gustavsson2006}; opening an entirely new research avenue. Other groups extended the methodology to include bidirectional transitions by including a double quantum dot in the detection process \cite{Fujisawa2006}, and demonstrated that bimodal current distributions are possible \cite{Fricke2007}. Single-electron detection was also used to measure electron-electron interference in a double quantum dot \cite{Gustavsson2008,Gustavsson2009}, the FCS of superconducting junctions \cite{Maisi2014}, the FCS in the transient regime \cite{Flindt2009}, non-Gaussian fluctuations \cite{Gershon2008}, and finite-frequency FCS \cite{Ubbelohde2012}. Sukhorukov et al. \cite{Sukhorukov2007} analysed the effect that single-electron detection has on transport, showing that the FCS is altered by the back-action of the QPC on the QD. 

Although most experiments have operated at ultra-low temperatures, usually in the mK range, recently a group has manged to use the optical blinking of a nearby semiconductor nanocrystal to make room temperature measurements on a carbon nanotube \cite{Zbydniewska2015}. A similar method, using resonance fluorescence, has analysed spin dynamics in a quantum dot \cite{Kurzmann2018}. Single-electron counting methods are unfortunately, however, still restricted to extremely low currents: approximately $10^{3}$ electrons per second.}

Likewise, theoretical calculations of the FCS initially faced great difficulty because, although the noise can be explicitly written as {{}
\begin{align}
\mathcal{S}(\omega) & = \int_{-\infty}^{\infty} dt\:e^{i\omega t} \left\langle \left[\delta\hat{I}(t),\delta\hat{I}(0)\right]_{+}\right\rangle, \label{noise definition}
\end{align}
} where $\delta\hat{I}(t) = \hat{I}(t) - \langle I \rangle$ and $\delta\hat{I}(t) = \hat{I}(0) - \langle I \rangle$, directly calculating higher-order cumulants is non-trivial, since quantum mechanical current operators at different times $\hat{I}(t)$ and $\hat{I}(t')$ in general do not commute. The time-ordering problem was first solved by Levitov and Lesovik \cite{Levitov1996} in a scattering theory framework via the explicit inclusion of a measuring device in the theoretical setup, which has since been applied extended to a general quantum mechanical variable \cite{Nazarov2003} and extended to various transport scenarios \cite{Lesovik2006,Hassler2008}. Non-equilibrium Green's functions have proved indispensable in calculating transient FCS for both electron \cite{Tang2014a} and electron energy \cite{Yu2016} current, as well as the FCS of junctions with electron-phonon \cite{Park2011, Agarwalla2015, Wang2011,Ueda2017} and electron-electron \cite{Ridley2018} interactions; although for strong interactions quantum master equations are generally used. Master equations have been used to calculate the FCS in various systems \cite{nazarov-book}, such as multi-level quantum dots \cite{Belzig2005}, systems with cotunneling effects \cite{Kaasbjerg2015, Emary2009, Utsumi2006, Leijnse2008}, systems with strong electron-phonon interactions \cite{Simine2012}, and systems experiencing the Coulomb blockade \cite{Bagrets2003}. {{} The FCS of non-Markovian transport are calculated from a general master equation, such as that in Eq.\eqref{standard non-Markovian ME}. Such work has shown, for example, that non-Markovian behavior enters the higher-order cumulants via cotunneling effects \cite{Braggio2006}, and is also present in the FCS of a dissipative double quantum dot\cite{Flindt2008,Flindt2010}. The FCS are closely related to fluctuation theorems: those relations that describe fluctuations in far-from-equilibrium quantum systems \cite{Esposito2009}.}

 In the Markovian master equation framework, current cumulants are expressed in terms of cumulants of the distribution of total transferred charge $P(n,t)$. The first and second cumulants, for example, are just the average $\langle n(t) \rangle$ and the variance $\langle\langle n(t)^{2} \rangle\rangle = \langle \left(n(t) - \langle n(t) \rangle\right)^{2}\rangle$.

{{} Assuming that the distribution of transferred charge, $P(n,t)$, has an exponential asymptotic behavior \cite{Budini2011,Touchette2009}
\begin{align}
\lim_{t\rightarrow\infty} P(n,t) & \approx \exp\left[-t\phi(\frac{n}{t})\right], \label{asymptotic behavior}
\end{align}
where $\phi(\frac{n}{t}) $ is the rate function of large deviation process,
then the cumulants of transferred charge in the long time limit $t\rightarrow\infty$ grow linearly according to the asymptotic growth rates $\langle\langle I^{k} \rangle\rangle$:

\begin{align}
\langle\langle n(t) ^{k} \rangle\rangle & \approx \langle\langle I^{k} \rangle\rangle t. \label{cumulant scaling time}
\end{align}

The approximation in Eq.\eqref{asymptotic behavior} implies that $P(n,t)$ satisfies a large deviation principle \cite{Touchette2009}. In essence, measurements of $n$ in the time interval $[0,t]$ will focus around the average $\langle n(t) \rangle = \min_{n} \left\{\phi(\frac{n}{t})\right\}$ and large deviations away from this value are exponentially suppressed. The large deviation principle is satisfied for a wide variety of physical systems, including the ones we consider here, and it forms the basis of the connections between the fluctuation statistics we present. It should be noted, however, that not all transport scenarios display the scaling in Eq.\eqref{cumulant scaling time}; Karzig and von Oppen, for example, have shown that the current cumulants in a chain of quantum dots do not scale linearly in time while undergoing a phase transition \cite{Karzig2010}.}

From Eq.\eqref{cumulant scaling time}, the current cumulants are related to the charge cumulants via 

\begin{align}
\lim_{t\rightarrow\infty}\langle\langle I(t) ^{k} \rangle\rangle & = e^{k}\frac{d}{dt}\langle\langle n(t)^{k} \rangle\rangle \\
& \approx \langle\langle I^{k} \rangle\rangle,
\end{align}
since we set $e = 1$. This is the reason we chose $\langle\langle I^{k} \rangle\rangle$ as notation for the asymptotic growth rates in Eq.\eqref{cumulant scaling time}. Clearly, the long-time limit is crucial to an easy relationship between current cumulants and cumulants of transferred charge. As we will see, it also arises from the master equation theory in order to keep computational simplicity. Calculating cumulants in the long-time limit, however, also restricts their Fourier transforms to the zero-frequency limit. We will demonstrate with the noise, as defined in Eq.\eqref{noise definition}, and the second cumulant of $P(n,t)$ \cite{Nazarov2003}.

Direct differentiation of the second cumulant gives
\begin{align}
\frac{d}{dt} \langle \langle n(t)^2 \rangle \rangle & = \left \langle \left[ n(t), \frac{d}{dt} n(t) \right]_{+} \right\rangle - 2 \langle n(t) \rangle \langle  \frac{d}{dt} n(t) \rangle.
\end{align}
By noting that
\begin{align}
n(t) & = \int_0^t dt_1 \hat I(t_1),
\end{align}
we then get 
\begin{align}
\frac{d}{dt} \langle \langle n(t)^2 \rangle \rangle & = \int_0^t dt_1 \left\langle \left[\delta \hat{I}(t_1) , \delta \hat{I}(t) \right]_{+}\right\rangle,
\end{align}
which already bears similarity to Eq.\eqref{noise definition}. Changing the integration variable from $t_1$ to relative time $\tau=t_1-t$ gives
\begin{align}
\frac{d}{dt} \langle \langle n(t)^2 \rangle \rangle & = \int_{-t}^0 d \tau \left\langle \left[\delta \hat{I}(t+\tau) , \delta \hat{I}(t) \right]_{+}\right\rangle.
\end{align}
In the limit $t\rightarrow\infty$, the measurement time $t$ is greater than the characteristic current-current correlation time $\tau$ and we can replace the integration limits to get the so-called MacDonald formula \cite{Macdonald1949}:
\begin{align}
\frac{d}{dt} \langle \langle n(t)^2 \rangle \rangle & =\frac{1}{2} \int_{-\infty}^{\infty} d \tau  \left\langle \left[\delta \hat{I}(t+\tau) , \delta \hat{I}(t) \right]_{+}\right\rangle. \label{Macdonald formula}
\end{align}
In the stationary state, Eq.\eqref{Macdonald formula} depends only on the time delay $\tau$ and, comparing with Eq.\eqref{noise definition}, therefore reduces to half the zero-frequency noise power
\begin{align}
\lim_{t\rightarrow\infty}\frac{d}{dt} \langle \langle n(t)^2 \rangle \rangle & \approx \langle\langle I^{2} \rangle\rangle \\
& = \frac{1}{2} \mathcal{S}(0);
\end{align}
so we see that the long-time limit necessarily restricts cumulants to the zero-frequency power regime, where they potentially miss important short-time physics. {{} Counting statistics at finite times, or rather finite frequencies, remain an active theoretical \cite{Galaktionov2003,Nagaev2004,Pilgram2004,Galaktionov2011,Marcos2010,Salo2006,Benito2016,Stegmann2017} and experimental \cite{Ubbelohde2012} research area; non-Poissonian behavior of higher-order cumulants, for example, has been shown to depend on frequency \cite{Emary2007}. Time-dependent current cumulants are able to identify short-time correlations between electrons \cite{Stegmann2016,Stegmann2018}; in fact it has been proposed that higher-order factorial cumulants can be used as a detection technique for electron-electron interactions \cite{Kambly2011,Kambly2013,Stegmann2015}. Despite the success of finite-frequency FCS, we restrict our analysis to the zero-frequency current cumulants as they are numerically easier to compute yet still provide valuable transport information.} 

Scaling the zero-frequency noise by the average current, for example, yields a quantity known as the Fano factor:
\begin{align}
F & = \frac{\mathcal{S}(0)}{2\langle I \rangle} \\
& = \frac{\langle\langle n ^{(2)}\rangle\rangle}{\langle n \rangle}. \label{FF n definition}
\end{align}
Since the variance and mean of a Poisson process are equal, the Fano factor can be used to characterise current distributions as either sub-Poissonian ($F<1$), Poissonian ($F=1$), or super-Poissonian ($F>1$), and consequently identify the transport effects causing this behavior. Indeed, super-Poissonian noise is caused by a host of physical effects, such as the dynamical channel blockade \cite{Belzig2005, Cottet2004}, asymmetric couplings \cite{Carmi2012}, avalanching electrons \cite{Koch2005}, telegraphic switching \cite{Ptaszynski2017,Ptaszynski2017a,Thielmann2005a,Kaasbjerg2015}, and negative differential resistance \cite{Thielmann2005}.

\subsection{Practical calculations}

In the introduction to this section, we saw that in the long-time limit FCS are recast in terms of the distribution of transferred charge $P(n,t)$. Fortunately, in the framework of master equations, there is an easy path to calculating cumulants of this distribution \cite{Bagrets2003}. First, $P(n,t)$ is related to the density matrix via the trace:

\begin{align}
P(n,t) & = \left(\mathbf{I},\mathbf{P}(n,t)\right),
\end{align}
where the notation $\left(\mathbf{I},\mathbf{A}\right)$ represents the inner product of vector $\mathbf{A}$, with dimension $m\times1$, with $\mathbf{I}$: a vector of ones with dimension $1\times m$.

From Eq.\eqref{Fourier transformed probability vector}, one can see that 

\begin{align}
\left(\mathbf{I},\mathbf{P}(\chi,t)\right) & = \sum_{n=0}^{\infty} e^{in\chi} P(n,t) \label{Chi inner product}.
\end{align}
If we take successive derivatives of Eq.\eqref{Chi inner product} with respect to $\chi$ and set $\chi = 0$, the right-hand side will produce the moments of transferred charge, and therefore the current:

\begin{align}
\langle I^{k}\rangle & = \frac{d}{dt} (-i)^{k} \frac{\partial^{k}}{\partial\chi^{k}} \left(\mathbf{I},\mathbf{P}(\chi,t)\right)\Bigg|_{\chi=0} \\
& = \frac{d}{dt} (-i)^{k} \frac{\partial^{k}}{\partial\chi^{k}} \left(\mathbf{I},e^{\mathbf{L}(\chi)t}\bar{\mathbf{P}}\right)\Bigg|_{\chi=0} \label{full current moment definition},
\end{align}
using Eq.\eqref{Chi dependent ME solution} and assuming all measurements are performed in the stationary state. The moment generating function of $P(n,t)$ is then $M(\chi,t) = \left(\mathbf{I},e^{\mathbf{L}(\chi)t}\bar{\mathbf{P}}\right)$; taking its natural logarithm one finds the cumulant generating function $K(\chi,t)$:

\begin{align}
K(\chi,t) & = \ln\left(\mathbf{I},e^{\mathbf{L}(\chi)t}\bar{\mathbf{P}}\right). \label{cumulant generating function current}
\end{align}

Eq.\eqref{cumulant generating function current} is, in general, too difficult to evaluate exactly. Bagrets and Nazarov \cite{Bagrets2003, Nazarov2003}, making the same assumption about the asymptotic behavior of $P(n,t)$ as in Eq.\eqref{asymptotic behavior}, have shown that as $t\rightarrow\infty$ the cumulant generating function can be approximated as 

\begin{align}
\lim_{t\rightarrow\infty} K(\chi,t) = t \Lambda_{max}, \label{eigenvalue simplification}
\end{align}
where $\Lambda_{max}$ is the eigenvalue of $M(\chi,t)$ with the largest real part. {{} Analytical calculations of eigenvalue in Eq.\eqref{eigenvalue simplification}  can be difficult  for systems with large Liouvillian matrix. The common approach then is to expand $\Lambda_{max}(\chi)$ in $\chi$ using Rayleigh-Schr\"{o}dinger perturbation theory \cite{Flindt2005,Flindt2008,Flindt2010}}. 

Cumulants of the current distribution are now
\begin{align}
\langle\langle I^{k}\rangle\rangle & = \frac{d}{dt} (-i)^{k} \frac{\partial^{k}}{\partial\chi^{k}} t\Lambda_{max}\Bigg|_{\chi=0} \\
& =  (-i)^{k} \frac{\partial^{k}}{\partial\chi^{k}} \Lambda_{max}\Bigg|_{\chi=0}. \label{full current cumulant definition}
\end{align}

\subsection{Equilibrated Phonons}

For equilibrated phonons, we can calculate the FCS analytically. Considering the $\chi$-dependent Liouvillian defined by Eq.\eqref{Full Liouvillian Chi}, Eq.\eqref{equilibrated Lo}, and Eq.\eqref{equilibrated jump operators}

\begin{align}
\mathbf{L}(\chi) & = \left[\begin{array}{cc} -T_{10} & T_{01}(\chi) \\ T_{10}(\chi) & -T_{01} \end{array}\right],
\end{align}
where $T_{10}(\chi) = T^{S}_{01} + T^{D}_{01}e^{i\chi} \text{ and }T^{S}_{10} + T^{D}_{10}e^{-i\chi}$, one can immediately read off the eigenvalue with the largest real part:

\begin{widetext}
\begin{align}
t\Lambda_{max}(\chi) & = \frac{t}{2}\left[-(T_{01} + T_{10}) + \sqrt{(T_{01} + T_{10})^{2} - 4[T_{01}T_{10} - T_{10}(\chi)T_{01}(\chi)]}\right].
\end{align}
\end{widetext}

Treating $t\Lambda_{max}(\chi)$ as the moment generating function and using Eq.\eqref{full current moment definition}, the average current is
\begin{align}
\langle I \rangle & = (-i)\frac{\partial}{\partial \chi} \Lambda_{max}(\chi) \Bigg|_{\chi=0} \\
& = \frac{T_{10}^{S}T_{01}^{D} - T_{01}^{S}T_{10}^{D}}{T_{10} + T_{01}};
\end{align}
the variance is
\begin{widetext}
\begin{align}
\langle\langle I^{2} \rangle\rangle & = (-i)^{2}\frac{\partial^{2}}{\partial \chi^{2}} \Lambda_{max}(\chi) \Bigg|_{\chi=0} \\
& = \frac{1}{(T_{01}+T_{10})^{3}}\left[(T_{01}^{S}T_{10}^{D} + T_{10}^{S}T_{01}^{D})(T_{01}+T_{10})^{2} - 2(T_{10}^{S}T_{01}^{D} - T_{01}^{S}T_{10}^{D})^{2}\right];
\end{align}
\end{widetext}
and the Fano factor is
\begin{widetext}
\begin{align}
F & = \frac{(T_{01}^{S}T_{10}^{D} + T_{10}^{S}T_{01}^{D})(T_{01}+T_{10})^{2} - 2(T_{10}^{S}T_{01}^{D} - T_{01}^{S}T_{10}^{D})^{2}}{(T_{10}^{S}T_{01}^{D} - T_{01}^{S}T_{10}^{D})(T_{01}+T_{10})^{2}}. \label{FCS exact Fano}
\end{align}
\end{widetext}

\section{Fluctuating-time statistics} \label{Fluctuating-time statistics Section}

Fluctuating-time statistics provide an alternative view of the transport. Consider the single-electron detection experiments outlined in the introduction of Section (\ref{Fixed-time statistics Section}). One could just as easily use the QPC to record the time between successive charging and discharging processes on the molecule. Repeated measurements could then be used to generate the WTD, $w(t_{a},t_{b})$: the conditional probability density that, given an electron tunneling event occured at time $t_{a}$, the next electron tunneling event occurs at time $t_{b}$ \cite{VanKampen1981}. This measurement scheme does not have to be restricted to electron tunneling events; one could measure a WTD in a similar manner for any situation where physical events occur at specific but random points in time. {{} Indeed, waiting times have a broad history in multiple disciplines, such as queueing theory \cite{Chowdhury2011} in mathematics, reaction kinetics \cite{Saha2011} in chemistry, and quantum optics \cite{Srinivas2010,Budini2010,Cao2006,Osadko2011} in physics.} They are, however, a relatively recent addition to mesoscopic transport; the first formalism was outlined by Brandes in 2008 \cite{Brandes2008}.

Since their introduction, waiting times have been calculated for a wide variety of mesoscopic transport scenarios. {{} Scattering theory, for example, has been used to calculate waiting times in superconducting junctions \cite{Albert2016,Chevallier2016,Mi2018}, periodically driven transport \cite{Albert2011,Albert2014}, and coherent conductors \cite{Dasenbrook2015,Haack2014,Albert2012}. As with the FCS, waiting times in the transient regime are generally calculated via the non-equilibrium Green's functions method \cite{Tang2014a}, which has been used to analyse the role of spin \cite{Tang2014a,Tang2018} and molecular vibrations \cite {SeoaneSouto2015} in electron transport.} Alongside these two methods, substantial research has followed Brandes' original formalism and calculated waiting times from quantum master equations. Walldorf et al. \cite{Walldorf2018} and Rajabi et al. \cite{Rajabi2013}, for example, have both used master equations to explore the relationship between waiting times and Cooper pair emission in superconducting junctions, while Potanina and Flindt \cite{Potanina2017} have investigated periodically driven electron transport. Waiting times have also been calculated for double-quantum dots \cite{Welack2009,Brandes2008,Ptaszynski2017a,Ptaszynski2017}, quantum dot spin-valves \cite{Sothmann2014}, non-Markovian transport \cite{Thomas2013}. Further work in the master equation framework has shown that waiting times can identify transport through HOMO and LUMO levels in a single resonant level \cite{Rudge2016a}, analyse electron-electron interactions in the  sequential \cite{Rudge2016,Ptaszynski2017,Brandes2008} and cotunneling \cite{Rudge2018} regimes, and analyse electron-phonon interactions \cite{Kosov2016,Kosov2017b,Kosov2018a,Koch2006}. 

Unlike FCS, which in quantum master equations is generally restricted to the long time regime, waiting times have no such restriction and can thus provide insight into interesting physics on short timescales. Phenomena such as inelastic interactions \cite{Kosov2017b,Rudge2018}, quantum coherence \cite{Dasenbrook2015,Albert2014}, fermionic statistics \cite{Ptaszynski2017,Albert2012}, spin-polarised leads \cite{Ptaszynski2017a}, and superconducting junctions \cite{Albert2016,Chevallier2016} have all been shown to produce temporal correlations observable from the WTD. Experimentally, however, waiting times do have several drawbacks. Since measurements rely on single-electron detection, as with the FCS, all currents must be small: in the order of $10^{3}$ counts per second. Furthermore, if one wishes to analyse purely quantum processes in the transport, such as cotunneling, then direct measurement via a QPC is impossible as it will destroy any coherence \cite{Kaasbjerg2015}. An interesting experimental approach, in which the WTD is extracted directly from low-order current correlations via theoretical post-processing using continuous matrix product state tomography \cite{Haack2015}, has recently been proposed and could possibly overcome this limitation. The experimental setup of real-time single electron detection is also restricted to large bias voltages, so that the current is unidirectional. We will see that this limit also applies in theoretical calculations as well; we are restricted to calculating the WTD for successive tunnelings to and from the drain separately. Considering that bidirectional transitions play an important role outside of the large bias limit and in many physical systems \cite{Fujisawa2006}, it is imperative to have a fluctuating-time statistic capable of incorporating them.  From a theoretical perspective, we could work with bidirectional transitions if there was a fluctuation statistic for the time between the $\text{\it{total}}$ number of forward and backward tunnelings.

Recently, it has been demonstrated that the first-passage time is such a statistic. In the context of mesoscopic electron transport, we define the FPTD $F(n|t_{0},t_{0}+\tau)$ as the probability density that, given an initial tunneling event to the drain at time $t_{0}$, the next time the total number of transferred electrons reaches $n$ is after a time delay $\tau$. Since $n$ is the difference between the sum of forward and backward transitions, it naturally incorporates bidirectional transport. First-passage times, in the form that we will use, were initially developed to describe fluctuations of entropic variables in the stationary state of Markovian systems \cite{Bauer2014,Roldan2015,Neri2017,Garrahan2017,Saito2016,Ptaszynski2018}; they have since been used to theoretically and experimentally verify fluctuation relations \cite{Singh2018}. We note, also, that Ridley et al. have recently calculated FPTDs using the inchworm quantum Monte Carlo method and found queuing effects arising from the Coulomb repulsion \cite{Ridley2018}. The notation used in this review was first outlined by Saito and Dhar \cite{Saito2016}, with significant contributions from Ptaszynski \cite{Ptaszynski2018}. The FPTD has since been useful in analysing systems where bidirectional transitions are unavoidable \cite{Rudge2019}.

In this section we will first outline general relationships between time statistics for fluctuating variables, before detailing methods for calculating the WTD and FPTD from Markovian master equations. We will conclude with an example using transport through the Holstein model for equilibrated phonons, and a discussion on renewal theory.

\subsection{Outline}

We first define several important probability distributions associated with individual electron tunneling events and establish various useful relations between these distributions.  We will closely follow van Kampen's discussion of stochastic time distributions\cite{VanKampen1981}: 
\\
\\
 $F(t_{a},t_{b})$ -- the probability {\it density} that, given that the recording starts at time $t_{a}$, the first electron tunneling event is detected at time $t_{b}$. Therefore, $F(t_{a},t_{b}) d t_{b}$ is the probability to first detect an electron at time interval $(t_{b},t_{b}+dt_{b})$ if the recording of the events starts at $t_{a}$; and 
 \\
 \\
 $\Pi(t_{a},t_{b})$ -- the probability  that no electron detection occurs in the interval $(t_{a},t_{b})${{}, which has recently been named the idle-time probability \cite{Albert2012,Haack2014}.}
 \\
 \\
These two probability distributions are connected by the self-evident integral relation
\begin{align}
\int_{t_{a}}^{t_{b}} d t \; F(t_{a},t)  & =  1- \Pi(t_{a},t_{b}).
\end{align}
Differentiating this relation with respect to $t_{b}$  and $t_{a}$ gives, respectively,
\begin{align}
F(t_{a},t_{b}) & = - \frac{\partial}{\partial t_{b}} \Pi(t_{a},t_{b}) \\
& = \frac{\partial}{\partial t_{a}} \Pi(t_{a},t_{b}).
\end{align}

There are three more important probability distributions:
\\
\\
$p(t)$ -- the probability density to detect an electron at time $t$, which, based on physical reasoning, satisfies $p(t_{a}) =F(t_{a},t_{a})$. An electron tunneling event occuring in the interval $[t,t+dt]$ has associated probability $p(t)dt$;
\\
\\
$p(t_{a},t_{b})$ -- the {\it joint} probability density to detect an electron at time $t_{a}$ and to next detect an electron at time $t_{b}$. The joint probability for an electron tunneling event to occur in $[t_{a},t_{a}+dt_{a}]$ and the next electron tunneling event to occur in $[t_{b},t_{b}+dt_{b}]$ is then $p(t_{a},t_{b})dt_{a}dt_{b}$; and
\\
\\
$w(t_{a},t_{b})$ -- the {\it conditional} probability density that, given an electron tunneling occurred at time $t_{a}$, the next electron tunneling event occured at $t_{b}$: the WTD. The expression $w(t_{a},t_{b})dt_{b}$ is thus the conditional probability that, given an electron tunneling occurred at time $t_{a}$, the next electron tunneling event occured in the time interval $[t_{b},t_{b}+dt_{b}]$.

Using the standard relation between joint and conditional probabilities,
\begin{align}
p(t_{a},t_{b}) dt_{a} dt_{b} & = p(t_{a}) dt_{a} \; w(t_{a},t_{b}) dt_{b},
\end{align}
we get a simple expression for the WTD in terms of the joint probability density:
\begin{align}
p(t_{a},t_{b}) &  = p(t_{a}) \; w(t_{a},t_{b}) \\
w(t_{a},t_{b}) &  = \frac{p(t_{a},t_{b})}{p(t_{a})}.
\end{align}

Based on these definitions, the probability density for an electron tunneling to occur at $t_{b}$, irrespective to any prior tunnelings before measurement started at $t_{a}$, is also 

\begin{align}
\int_{-\infty}^{t_{a}} dt \: p(t, t_{b}) & = F(t_{a},t_{b}).
\end{align}
Differentiating with respect to $t_{a}$ we get
\begin{align}
p(t_{a},t_{b}) & = \frac{\partial}{\partial t_{a}} F(t_{a},t_{b}).
\end{align}
Hence, we have the following relations between the WTD and idle-time probability
\begin{align}
w(t_{a},t_{b}) & =- \frac{1}{p(t_{a})} \frac{\partial^2}{\partial t_{a} \partial t_{b}} \Pi(t_{a},t_{b}), \text{ and} \\
w(t_{a},t_{b}) & = \frac{1}{p(t_{a})} \frac{\partial^2}{\partial t_{a}^2} \Pi(t_{a},t_{b}).
\end{align}

As in the previous section we perform all calculations in the stationary non-equilibrium state, where all two-time distributions now depend only on the relative time $\tau=t_{b}-t_{a}$. Using $\frac{\partial}{\partial \tau} = \frac{\partial}{\partial t_{b}} = -\frac{\partial}{\partial t_{a}} $ we get
\begin{equation}
w(\tau) = \frac{1}{p} \frac{\partial^2}{\partial \tau^2}  \Pi(\tau), \label{WTD idle time definition}
\end{equation}
with $p$ now also computed from the idle time probability:
\begin{equation}
p =\left.  -\frac{\partial}{\partial \tau}  \Pi(\tau) \right|_{\tau=0}. \label{initial probability}
\end{equation}

\subsection{Waiting time distribution}

For a generic quantum system described by the Markovian master equation in Eq.\eqref{standard ME}, one can intuitively generate the distribution of waiting times between successive electron tunneling events. For simplicity, let us assume that we are examining the waiting time $\tau$ between two electron tunneling events of the same type, described by jump operator $\mathbf{J}$. The conditional probability density that, given that tunneling event of type $\mathbf{J}$ occurs at some time in the stationary state, the next tunneling event of type $\mathbf{J}$ will occur after a delay $\tau$ is 

\begin{align}
w(\tau) & = \frac{\left(\mathbf{I},\mathbf{J}e^{\mathbf{L}_{0}\tau}\mathbf{J}\bar{\mathbf{P}}\right)}{\left(\mathbf{I},\mathbf{J}\bar{\mathbf{P}}\right)}, \label{brandes definition}
\end{align}
where $\mathbf{L}_{0} = \mathbf{L} - \mathbf{J}$ is the Liouvillian with $\mathbf{J}$ removed. 

Let us physically examine the top line of Eq.\eqref{brandes definition}. The system starts in the stationary state $\bar{\mathbf{P}}$ and undergoes quantum jump $\mathbf{J}$, after which it evolves for a time according to $\mathbf{L}_{0}$, during which no jump of type $\mathbf{J}$ takes place, until another quantum jump $\mathbf{J}$ occurs after a time $\tau$. Summing the resulting probability vector, which is equivalent of computing the inner product with vector $\mathbf I$, thus gives the joint probability for two successive quantum jumps of type $\mathbf{J}$ to occur  in the stationary state at times separated by a delay $\tau$. The bottom line is just the probability for quantum jump $\mathbf{J}$ to occur at any time during the stationary state; so that, together, the top and bottom lines of Eq.\eqref{brandes definition} denote the conditional probability that, given an initial quantum jump of type $\mathbf{J}$ in the stationary state, the next quantum jump of type $\mathbf{J}$ will occur after a waiting time $\tau$, which is Brandes' original definition \cite{Brandes2008}.

Since we aim to relate time statistics to current statistics, we will focus on $w_{F}(\tau)$, the distribution of waiting times between tunnelings from the molecule to the drain, which are contained in the forward current operator $\mathbf{J}_{F}$, and $w_{B}(\tau)$, the distribution of waiting times between tunnelings from the drain to the molecule, which are contained in the backward current operator $\mathbf{J}_{B}$.

If one assumes that the transport is unidirectional, such that $\mathbf{L}(\chi) = \mathbf{L}_{0} + \mathbf{J}_{F}e^{i\chi}$, then one can derive an equivalent expression for the WTD from the $n$-resolved master equation and the idle-time probability $\Pi(t)$, since in this limit

\begin{align}
\Pi(t) & = P(0,t).
\label{fcs-16}
\end{align}
Again, this follows from physical intuition; $\Pi(\tau)$ is the probability that at time $\tau$ no tunneling event has occurred, which for unidirectional transport is $P(0,t)$ since $n \geq 0$. The probability $P(0,t)$ is obtained from the cumulant generating function, defined in Eq.\eqref{cumulant generating function current}:

\begin{align}
M(\chi,t) & = e^{K(\chi,t)} \label{MGF from CGF} \\
& = \sum_{n=0}^{\infty} e^{i n \chi} P(n,t) \\
& = P(0,t) + \sum_{n=1}^{\infty} e^{i n \chi} P(n,t),
\end{align}
where all the terms in the summation for which $n<0$ are excluded since the transport is unidirectional. All the terms inside the summation disappear in the limit $\chi \rightarrow i \infty$, {{} so that \cite{Thomas2013,Dasenbrook2016}}

\begin{align}
P(0,t) & = \lim_{\chi \rightarrow i \infty} M(\chi,t).
\end{align}
which, combined with Eq.\eqref{cumulant generating function current}, yields

\begin{align}
\Pi(\tau) & = \lim_{\chi \rightarrow i\infty} \left(\mathbf{I},e^{\mathbf{L}(\chi)t}\bar{\mathbf{P}}\right). \label{ME idle time probability}
\end{align}
Here we immediately see the necessity for excluding all terms for which $n<0$; if included each would be accompanied by a factor of $e^{-i\chi}$, which would diverge in the limit $\chi \rightarrow i\infty$.

From here, we proceed using Eq.\eqref{WTD idle time definition} alongside the definition of $\Pi(\tau)$ in Eq.\eqref{ME idle time probability}:
\begin{widetext}
\begin{align}
w_{F}(\tau) & = -\lim_{\chi\rightarrow i\infty}\frac{(\mathbf{I},\mathbf{L}(\chi)e^{\mathbf{L}(\chi)\tau}\mathbf{L}(\chi)\bar{\mathbf{P}})}{(\mathbf{I},\mathbf{L}(\chi)\bar{\mathbf{P}})}. \\
& = -\lim_{\chi\rightarrow i\infty}\frac{(\mathbf{I},(\mathbf{L}_{0}+\mathbf{J}_{F}e^{i\chi})e^{(\mathbf{L}_{0}+\mathbf{J}_{F}e^{i\chi})\tau}(\mathbf{L}_{0}+\mathbf{J}_{F}e^{i\chi})\bar{\mathbf{P}})}{(\mathbf{I},(\mathbf{L}_{0}+\mathbf{J}_{F}e^{i\chi})\bar{\mathbf{P}})} \\
& = -\frac{(\mathbf{I},\mathbf{L}_{0}e^{\mathbf{L}_{0}\tau}\mathbf{L}_{0}\bar{\mathbf{P}})}{(\mathbf{I},\mathbf{L}_{0}\bar{\mathbf{P}})}. \label{n resolved WTD definition}
\end{align} 
\end{widetext}
At this point we use the splitting $\mathbf{L}_{0} = \mathbf{L}(0) - \mathbf{J}_{F}$, the definition of the stationary state $\mathbf{L}(0)\bar{\mathbf{P}}=0$, along with the easily verifiable relation, between any secular Liouvillian $\mathbf{L}(0)$ and any vector $\mathbf{A}$, that $\left(\mathbf{I},\mathbf{L}(0)\mathbf{A}\right)=0$, to obtain
\begin{align}
w_{F}(\tau) & = \frac{(\mathbf{I},\mathbf{J}_{F}e^{\mathbf{L}_{0}\tau}\mathbf{J}_{F}\bar{\mathbf{P}})}{(\mathbf{I},\mathbf{J}_{F}\bar{\mathbf{P}})}. \label{WTD definition n resolved matches brandes}
\end{align} 

Eq.\eqref{WTD definition n resolved matches brandes} is the same definition as that provided in Eq.\eqref{brandes definition}. The important distinction, however, is that while Eq.\eqref{brandes definition} can include bidirectional transitions in $\mathbf{L}_{0}$, from the very start of deriving Eq.\eqref{WTD definition n resolved matches brandes} we are forced to assume that the transport is unidirectional. One may of course use the same approach to derive a similar expression for $w_{B}(\tau)$ from the idle-time probability in the limit $\chi \rightarrow -i\infty$, but the assumption would then be that the transport is unidirectional in the $n \leq 0$ direction.

To resolve this backward tunneling divergence catastrophe, one might intuitively return to Eq.\eqref{Chi inner product} and Eq.\eqref{cumulant generating function current} and write the moment generating function as 
\begin{align}
M(\chi,t) & = P(0,t) + \sum_{n=1}^{\infty} e^{i n \chi} P(n,t) + \sum_{n=-1}^{-\infty} e^{i n \chi} P(n,t).
\end{align}
Integrating both sides from $0\rightarrow2\pi$ will thus eliminate all terms for which $n>0$ and $n<0$:
\begin{align}
P(0,t) & = \int^{2\pi}_{0} d\chi \: M(\chi,t). 
\end{align}

This method is nonetheless flawed, because for bidirectional transport $P(0,t) \neq \Pi(t)$. To see this, consider the physical definition of the idle-time probability $\Pi(\tau)$: the probability that {\it{no}} tunneling event occurs in the interval $[0,\tau]$. $P(0,t)$ is the probability that the jump number at time $t$ is $0$: that is, the probability that the sum of forward and backward transitions is zero. This does not preclude a tunneling event from occurring; indeed, there may have been any number of forward tunneling events in $[0,t]$, as long as there were also exactly the same number of backward tunneling events.

At this point it is natural to question the need for a WTD derived from the $n$-resolved master equation, as it is applicable to unidirecitonal transport only and one may obtain the same results from the definition in Eq.\eqref{brandes definition}. In some cases, however, the $n$-resolved master equation must be used to include all transitions that change the jump number $n$. Elastic cotunneling events, for example, do not change the state of the quantum system, and thus do not appear in the standard master equation, but contribute to the total current in the drain and must thus be included in $\mathbf{J}_{F}$ and $\mathbf{J}_{B}$. When one calculates waiting times including elastic cotunneling events, then, one must first define $\mathbf{L}(\chi)$ from the corresponding $n$-resolved master equation and then define $\mathbf{L}_{0} = \mathbf{L}(0) - \mathbf{J}_{F}$ from it \cite{Rudge2018,Rudge2019}. In fact, since waiting times in mesoscopic transport, defined via Markovian master equations, have largely been restricted to the infinite bias voltage regime, many authors define the WTD using the $n$-resolved approach \cite{Brandes2008,Thomas2013}.

Most waiting time analysis will be done via the cumulants, in particular the average waiting time $\langle\langle \tau \rangle\rangle = \langle \tau \rangle$ and the variance $\langle\langle \tau^{2} \rangle\rangle = \langle\tau^{2}\rangle - \langle\tau\rangle^{2}$. The Laplace transform of the WTD,
\begin{align}
\tilde{w}(z) & = \int^{\infty}_{0} d\tau \: e^{z\tau}w(\tau) \\
& = \frac{\left(\mathbf{I},\mathbf{J}\left[z - \mathbf{L}_{0}\right]^{-1}\mathbf{J}\bar{\mathbf{P}}\right)}{\left(\mathbf{I},\mathbf{J}\bar{\mathbf{P}}\right)},
\end{align}
conveniently defines a cumulant generating function:
\begin{align}
\langle\langle \tau^{k} \rangle\rangle = \left. (-1)^{k} \frac{d^{k}}{dz^{k}}\ln \tilde{w}(z)\right|_{z=0}. \label{WTD cumulants}
\end{align}

Scaling the second waiting time cumulant by the first yields a quantity known as the randomness parameter:
\begin{align}
R = \frac{\langle\langle \tau^{2} \rangle\rangle}{\langle\tau\rangle^{2}}, \label{randomness parameter}
\end{align}
which is commonly compared with the Fano factor.

We now have a comprehensive framework in which to calculate WTDs from Markovian master equations. Several problems remain, however. For those transport scenarios that do require an $n$-resolved master equation, how do we move beyond the unidirectional transport limit? Furthermore, we will see later that even if one is able to calculate $w_{F}(\tau)$ and $w_{B}(\tau)$ for bidirectional transport, it is not clear how to use them to obtain the total current distribution: one of the key interests in fluctuation analysis being the relations between fixed-time and fluctuating-time statistics. To resolve, we need a fluctuating-time distribution for the jump number $n$, which inherently includes both forward and backward transitions. Unfortunately, the idle-time probability offers no solution here either; for bidirectional transport $P(0,t)$ is the probability that $n=0$ at time $t$, which does not exclude the possibility that at some point before $t$ the jump number differed from zero, and hence is {\it{not}} the probability that $n$ did not change over $[0,t]$. Instead, we need a method to evaluate when the jump number {\it{first}} reaches $n$.

\subsection{First-passage time distribution}

{{} The quantity we seek is the FPTD $F(n|\tau)$: the probability density that the time delay until the jump number first reaches $n$ is $\tau$. Since $n$ is the sum of forward and backward transitions, the FPTD is naturally bidirectional. It is calculated from the $n$-resolved master equation: a relationship that was first outlined by Saito and Dhar \cite{Saito2016} and \cite{Ptaszynski2018}, whose work we closely follow in the derivations below.}

Considering the context of mesoscopic electron transport, it is intuitive to consider $F(n|t)$ as a trace over a first-passage time vector $\left(\mathbf{I},\mathbf{F}(n|t)\right)$, where each element $\left[\mathbf{F}(n|\tau)\right]_{l}$ is the probability density that the jump number reaches $n$ for the first time at $\tau$ and that the system is in state $l$ at this time. We next need to relate $\mathbf{F}(n|\tau)$ to the probability vector $\mathbf{P}(n,t)$, which means that the jump number $n$ cannot experience an overall change in the interval $\left[\tau,t+\tau\right]$. We therefore define $\mathbf{T}(0,t-\tau)$ as the matrix of conditional probability densities that, given at time given at time $\tau$ the jump number reaches $n$ for the first time, the jump number does not overall change in the time interval $[\tau,t]$. The probability vector $\mathbf{P}(n,t)$ is then the product of this conditional probability and the initial first-passage time probability, integrated over all possible first-passage times:

\begin{align}
\mathbf{P}(n,t) & = \int^{t}_{0} d\tau \: \mathbf{T}(0,t-\tau)\mathbf{F}(n,\tau). \label{Volterra equation}
\end{align}

In general, the element $\left[\mathbf{T}(n,t)\right]_{kl}$ is the conditional probability that, given the system is initially in state $l$ at time $t=0$, it will be in state $k$ at time $t$ and the jump number is $n$. One can see then, that $\mathbf{T}(n|t)$ defines a transition matrix moving the system from some arbitrary state at time $t=0$ to $\mathbf{P}(n,t)$:

\begin{align}
\mathbf{P}(n,t) & = \mathbf{T}(n|t)\mathbf{P}(0). \label{Transition matrix formula}
\end{align}

Equating Eq.\eqref{Volterra equation} and Eq.\eqref{Transition matrix formula} relates the first-passage time to the transition matrix:
\begin{align}
\mathbf{T}(n|t)\mathbf{P}(0) & = \int^{t}_{0} d\tau \: \mathbf{T}(0,t-\tau)\mathbf{F}(n,\tau). \label{Time definition of FPTD before rearrangement}
\end{align}
Since Eq.\eqref{Volterra equation} is a convolution, from here it is easier to work in Laplace space:

\begin{align}
\tilde{\mathbf{T}}(n|z)\mathbf{P}(0) & = \tilde{\mathbf{T}}(0,z)\tilde{\mathbf{F}}(n,z). \label{Laplace definition of FPTD before rearrangement}
\end{align}

Rearranging Eq.\eqref{Laplace definition of FPTD before rearrangement} yields the first-passage time distribution in Laplace space:
\begin{align}
\tilde{F}(n|z) & = \left(\mathbf{I},\tilde{\mathbf{T}}(0,z)^{-1}\tilde{\mathbf{T}}(n|z)\mathbf{P}(0)\right). \label{General FPTD Laplace}
\end{align}

All that remains now is to calculate $\mathbf{T}(n|t)$, which is obtained by comparing Eq.\eqref{Transition matrix formula} with the inverse Fourier transform of Eq.\eqref{Chi dependent ME}:

\begin{align}
\mathbf{T}(n|t) & = \frac{1}{2\pi}\int_{0}^{2\pi} d\chi e^{-in\chi} e^{\mathbf{L}(\chi)t}\mathbf{P}(0), \text{ and} \label{Transition matrix timespace definition} \\
\tilde{\mathbf{T}}(n|z) & = \frac{1}{2\pi}\int_{0}^{2\pi} d\chi e^{-in\chi} \left[z - \mathbf{L}(\chi)\right]^{-1}. \label{Transition matrix laplace definition}
\end{align}
Evaluating the contour integral in Eq.\eqref{Transition matrix laplace definition} must in general be done numerically, although for simple single reset systems multiple authors have calculated analytic results \cite{Ptaszynski2018,Saito2016}. 

The final step is to choose $\mathbf{P}(0)$ such that the FPTD relates directly to the WTD and FCS. For $n>0$, $\mathbf{P}(0)$ must therefore be the normalised probability vector after a forward jump in the stationary state:

\begin{align}
\mathbf{P}(0) & = \frac{\mathbf{J}_{F}\bar{\mathbf{P}}}{\left(\mathbf{I},\mathbf{J}_{F}\bar{\mathbf{P}}\right)}.
\end{align}
Similarly, the initial vector for $n<0$ is 
\begin{align}
\mathbf{P}(0) & = \frac{\mathbf{J}_{B}\bar{\mathbf{P}}}{\left(\mathbf{I},\mathbf{J}_{B}\bar{\mathbf{P}}\right)}.
\end{align}

With this definition, $F(n|\tau)$ is the probability density that the jump number first reaches $n$ at time $\tau$, given measurement started after an initial tunneling to the drain at $t=0$ in the stationary state, and $\tilde{F}(n|z)$ is its Laplace transform:

\begin{align}
\tilde{F}(n|z) & = \frac{\left(\mathbf{I},\tilde{\mathbf{T}}(0,z)^{-1}\tilde{\mathbf{T}}(n|z)\mathbf{J}_{F}\bar{\mathbf{P}}\right)}{\left(\mathbf{I},\mathbf{J}_{F}\bar{\mathbf{P}}\right)}. \label{FPTD Laplace Final}
\end{align}

The inverse Laplace transform yields an explicit equation for $F(n|\tau)$,
\begin{align}
F(n|\tau) & = \frac{1}{2\pi i}\lim_{R\rightarrow\infty}\int_{c-iR}^{c+iR}dz \: e^{z\tau}\frac{\left(\mathbf{I},\tilde{\mathbf{T}}(0,z)^{-1}\tilde{\mathbf{T}}(n|z)\mathbf{J}_{F}\bar{\mathbf{P}}\right)}{\left(\mathbf{I},\mathbf{J}_{F}\bar{\mathbf{P}}\right)}; \label{FPTD Time}
\end{align}
however, $F(n,z)$ is a convenient form for calculating cumulants of the FPTD:
\begin{align}
\langle\langle\tau^{k}_{n}\rangle\rangle^{*} & = (-1)^{k} \lim_{z\rightarrow 0^{+}}\left[\frac{d^{k}}{dz^{k}}\ln\tilde{F}(n|z)\right] \label{CGF FPTD}
\end{align}
Here, the notation $\langle\langle\tau^{k}_{n}\rangle\rangle^{*}$ translates to the $\text{k}^{th}$ cumulant of $F(n|\tau)$ and the limit $z\rightarrow 0^{+}$ is necessary since $\mathbf{L}(\chi)$ is singular for $\chi = \{0,2\pi\}$ \cite{Ptaszynski2018}.

As with the WTD and FCS, we focus on the first $\langle\tau_{n}\rangle^{*}$ and second $\langle\langle\tau^{2}_{n}\rangle\rangle^{*}$ cumulants, and their combination into the first-passage time randomness parameter:

\begin{align}
R^{*}_{n} & = \frac{\langle\langle\tau_{n}^{2}\rangle\rangle^{*}}{\left(\langle\tau_{n}\rangle^{*}\right)^{2}}. \label{RP FPTD definition}
\end{align}

\subsection{Equilibrated phonons}

\subsubsection{WTD}

For the simple case of equilibrated phonons the forward and backward tunneling WTDs, directly evaluated from Eq.\eqref{brandes definition}, are

\begin{align}
w_{F}(\tau) & = \frac{T_{10}T^{D}_{01}}{A}e^{-\frac{\tau}{2}(T_{01}+T_{10})}\left[e^{\frac{\tau}{2}A} - e^{-\frac{\tau}{2}A}\right] \text{ and} \label{Forward WTD definition}\\
w_{B}(\tau) & = \frac{T_{01}T^{D}_{10}}{B}e^{-\frac{\tau}{2}(T_{01}+T_{10})}\left[e^{\frac{\tau}{2}B} - e^{-\frac{\tau}{2}B}\right], \label{Backward WTD definition}
\end{align} 
where 
\begin{align}
A & = \sqrt{(T_{01} - T_{10})^{2} + 4T_{10}T^{S}_{01}} \text{ and}\\
B & = \sqrt{(T_{01} - T_{10})^{2} + 4T_{01}T^{S}_{10}}.
\end{align} 
Their Laplace transforms are 
\begin{align}
\tilde{w}_{F}(z) & = \frac{T_{10}T_{01}^{D}}{z^{2} + (T_{01}+T_{10})z + T_{01}T_{01}^{D}} \text{ and} \label{Forward WTD definition Laplace}\\
\tilde{w}_{B}(z) & =  \frac{T_{10}T_{01}^{D}}{z^{2} + (T_{01}+T_{10})z + T_{01}T_{10}^{D}}. \label{Backward WTD definition Laplace}
\end{align} 

From Eq.\eqref{WTD cumulants} the first and second cumulants are 
\begin{align}
\langle\tau\rangle_{F} & = \frac{T_{01}+T_{10}}{T_{10}T^{D}_{01}} \\
\langle\tau\rangle_{B} & = \frac{T_{01}+T_{10}}{T_{01}T^{D}_{10}}
\end{align}
and 
\begin{align}
\langle\langle\tau^{2}\rangle\rangle_{F} & = \frac{(T_{01})^{2}+T_{10}(T_{10}+2T_{01}^{S})}{(T_{10}T_{01}^{D})^{2}} \\
\langle\langle\tau^{2}\rangle\rangle_{B} & = \frac{(T_{10})^{2}+T_{01}(T_{01}+2T_{10}^{S})}{(T_{01}T_{10}^{D})^{2}},
\end{align}
respectively.

\subsubsection{FPTD} \label{Exact FPTD Cumulants}

The FPTD can also be written explicitly for equilibrated phonons. Indeed, multiple authors have done so for the equivalent scenario of a single-resonant level, which we outline here. For a full derivation, see Ref.[\onlinecite{Ptaszynski2018}]. We focus only on the case when $n>0$, for which the FPTD is

\begin{align}
\tilde{F}(n|z) & = \frac{[\tilde{\mathbf{T}}(n|z)]_{11}}{[\tilde{\mathbf{T}}(0|z)]_{11}}, \label{FPTD single reset definition}
\end{align}
where the simplification arises from the structure of $\mathbf{J}_{F}$. Using Eq.\eqref{Transition matrix laplace definition} the element $[\tilde{\mathbf{T}}(n|z)]_{11}$ is 
\begin{align}
[\tilde{\mathbf{T}}(n|z)]_{11} & = \frac{1}{2\pi}\int_{0}^{2\pi} d\chi e^{-in\chi} \left[ [z - \mathbf{L}(\chi)]^{-1}\right]_{11} \\
& = \frac{z+T_{01}}{2\pi}\int_{0}^{2\pi} \frac{d\chi}{e^{in\chi}}\frac{1}{\text{det}\left[z - \mathbf{L}(\chi)\right]}.
\end{align}

The determinant of $\left[z - \mathbf{L}(\chi)\right]$ defines an equation in $\chi$ with only $e^{i\chi}$, $e^{-i\chi}$, and $e^{0}$ terms. It can therefore be written as
\begin{multline}
[\tilde{\mathbf{T}}(n|z)]_{11} = \\
\frac{z+T_{01}}{2\pi}
 \int_{0}^{2\pi} \frac{d\chi}{e^{i(n-1)\chi}}\frac{1}{\left[e^{i\chi} - \lambda_{+}(z)\right]\left[e^{i\chi} - \lambda_{-}(z)\right]}. \label{T11 equation}
\end{multline}
The quantities $\lambda_{+}(z)$ and $\lambda_{-}(z)$ are the upper and lower roots of $\text{det}\left[z - \mathbf{L}(\chi)\right]$, respectively, and are known to satisfy $\lambda_{+}(z) > 1$ and $\lambda_{-}(z) < 1$ \cite{Ptaszynski2018}. Eq.\eqref{T11 equation} can then be solved directly using residue theory, as only one pole lies within the contour, or more easily as the $(n-1)^{\text{th}}$ term in the Laurent series expansion of $\left(\left[e^{i\chi} - \lambda_{+}(z)\right]\left[e^{i\chi} - \lambda_{-}(z)\right]\right)^{-1}$:
\begin{align}
[\tilde{\mathbf{T}}(n|z)]_{11} & = \frac{(z+T_{01})\lambda_{+}(z)}{\lambda_{-}(z) - \lambda_{+}(z)}.
\end{align}
From Eq.\eqref{FPTD single reset definition} the first-passage time distribution is then 
\begin{align}
\tilde{F}(n|z) & = \left[\lambda_{+}(z)\right]^{n} \label{EQ phonon FPTD} \\ 
& = [\tilde{F}(1|z)]^{n}. \label{Factored FPTD}
\end{align}
All that remains is evaluate $\lambda_{+}(z)$:
\begin{align}
\lambda_{+}(z) & = \frac{-b(z)}{2T_{10}^{S}T_{01}^{D}} + \frac{\sqrt{b(z)^{2}-4T_{10}^{S}T_{01}^{S}T_{10}^{D}T_{01}^{D}}}{2T_{10}^{S}T_{01}^{D}},
\end{align}
where 
\begin{align}
b(z) & = (z+T_{01})(z+T_{10}) - \left(T_{01}^{S}T_{10}^{S} + T_{01}^{D}T_{10}^{D}\right).
\end{align}

Eq.\eqref{Factored FPTD} demonstrates that for equilibrated phonons the FPTD can be factored and the cumulants are therefore linearly related:

\begin{align}
\langle\langle\tau^{k}_{n}\rangle\rangle^{*} & = \left.(-1)^{k} \lim_{z\rightarrow 0^{+}}\left[\frac{d^{k}}{dz^{k}}\ln\tilde{F}(1|z)^{n}\right]\right|_{z\rightarrow0} \\
& = n \langle\langle\tau^{k}_{1}\rangle\rangle^{*}. \label{FPTD cumulant relations}
\end{align}
The first and second cumulants of $F(1|\tau)$ are 
\begin{align}
\langle \tau_{1} \rangle^{*} & = \frac{T_{01}+T_{10}}{T_{10}^{S}T_{01}^{D} - T_{01}^{S}T_{10}^{D}}
\end{align}
and
\begin{widetext}
\begin{align}
\langle\langle \tau^{2}_{1} \rangle\rangle^{*} & = \frac{(T_{01}^{S}T_{10}^{D} + T_{10}^{S}T_{01}^{D})(T_{01}+T_{10})^{2} - 2(T_{10}^{S}T_{01}^{D} - T_{01}^{S}T_{10}^{D})}{(T_{10}^{S}T_{01}^{D} - T_{01}^{S}T_{10}^{D})^{3}},
\end{align}
respectively. Their combination, the FPTD randomness parameter, is 
\begin{align}
R^{*} & = \frac{(T_{01}^{S}T_{10}^{D} + T_{10}^{S}T_{01}^{D})(T_{01}+T_{10})^{2} - 2(T_{10}^{S}T_{01}^{D} - T_{01}^{S}T_{10}^{D})^{2}}{(T_{10}^{S}T_{01}^{D} - T_{01}^{S}T_{10}^{D})(T_{01}+T_{10})^{2}}. \label{FPTD exact RP}
\end{align}
\end{widetext}

\subsection{Renewal theory} \label{Higher-order time distributions}

Eq.\eqref{Factored FPTD} presents an interesting possibility for how multi-time distributions could be related to one another. It is, in fact, an example of a branch of analysis called renewal theory, which is based on the titular renewal assumption. For first-passage times, the factorisation in Eq.\eqref{Factored FPTD} is one expression of the renewal assumption. It can be written alternatively as \cite{Ptaszynski2018}

\begin{align}
F(n|\tau_{n} ; n'|\tau_{n'}) = F(n|\tau_{n}) \: F(n' -  n|\tau_{n'} - \tau_{n}), \label{FPTD renewal 2}
\end{align}
where $F(n|\tau_{n} ; n'|\tau_{n'})$ is the joint probability density that the jump number first reaches $n$ at time $\tau_{n}$ and first reaches $n'$ at time $\tau_{n'}$. We can write the renewal assumption similarly for waiting times:
\begin{align}
w_{2}(\tau,\tau') & = w(\tau)w(\tau'),
\label{wtd-renewal}
\end{align}
where $w_{2}(\tau,\tau')$ is the joint probability density that, given an initial tunneling, the system waits time $\tau$ until the next tunneling and then waits another time $\tau'$ for the tunneling after that. The renewal assumption therefore implies that successive waiting times are independently and identically distributed and the system state is ``renewed'' after each waiting time. The same logic follows for first-passage times.

If the renewal assumption is violated, then successive waiting times are no longer independent, temporal correlations emerge in quantum dynamics, and we see non-renewal behavior. As we will observe, non-renewal dynamics can emerge even under the Markovian assumption,  indicating that correlations between successive waiting times arise from the internal dynamics of the quantum system. {{} Although in mesoscopic electron transport, non-renewal statistics are a relatively new research premise  \cite{Dasenbrook2015,Ptaszynski2017,Kosov2017b,Ptaszynski2018,Rudge2019,Budini2011,Albert2011}, they have a long history in chemical physics, where they were used to describe single-molecule processes in spectroscopy\cite{Cao2006,Osadko2011,Budini2010, Witkoskie2006} and kinetics \cite{Saha2011,cao08}.}

Correlations between successive waiting times $\tau$ and $\tau'$ are described by the Pearson correlation coefficient:

\begin{align}
p  & = \frac{\langle\tau\tau'\rangle - \langle\tau\rangle^{2}}{\langle\langle\tau^{2}\rangle\rangle}. \label{Correlation coefficient wtd}
\end{align}
Here, $\langle \tau\tau'\rangle$ is the first moment of the second-order distribution $w_{2}(\tau,\tau')$ and we see a need to calculate moments of higher-order WTDs. 

Just as we can define probability distributions for time delays between two tunneling events, so can we also define distributions for multiple time delays between a series of tunneling events: the higher-order time distributions. Consider the second-order WTD \cite{Kosov2017b}:

\begin{align}
w_{2}(\tau,\tau') & = \frac{\left(\mathbf{I},\mathbf{J}e^{\mathbf{L}_{0}\tau'}\mathbf{J}e^{\mathbf{L}_{0}\tau}\mathbf{J}\bar{\mathbf{P}}\right)}{\left(\mathbf{I},\mathbf{J}\bar{\mathbf{P}}\right)}. \label{second order WTD}
\end{align}
The top line is the joint probability that the system starts in $\bar{\mathbf{P}}$ and undergoes quantum jump $\mathbf{J}$, after which it evolves according to $\mathbf{L}_{0}$, until another quantum jump of type $\mathbf{J}$ occurs after a time $\tau$, and the system again evolves according to $\mathbf{L}_{0}$ until the final quantum jump of type $\mathbf{J}$ occurs after $\text{\it{another}}$ waiting time $\tau'$. As usual, the bottom line provides the probability for quantum jump $\mathbf{J}$ to occur at any time during the stationary state, so that $w_{2}(\tau,\tau')$ is the probability density that three quantum jumps of type $\mathbf{J}$ will be separated by successive waiting times $\tau$ and $\tau'$, conditioned upon the probability density of the initial jump.

The first moment $\langle \tau\tau'\rangle$ is easily obtained via a moment generating function:
\begin{align}
\langle\tau\tau'\rangle & = \left.\frac{\partial}{\partial z}\frac{\partial}{\partial z'} \tilde{w}(z,z')\right|_{z=z'=0} \\
& = \frac{\left(\mathbf{I},\mathbf{J}\mathbf{L}_{0}^{-2}\mathbf{J}\mathbf{L}_{0}^{-2}\mathbf{J}\bar{\mathbf{P}}\right)}{\left(\mathbf{I},\mathbf{J}\bar{\mathbf{P}}\right)},
\end{align}
where 
\begin{align}
\tilde{w}(z,z') & = \int_{0}^{\infty} \int_{0}^{\infty} d\tau' \: d\tau \: e^{-(z\tau+z'\tau')}w(\tau,\tau').
\end{align}

The Pearson correlation coefficient between two subsequent first-passage times, $p^{*}$, follows the same definition as Eq.\eqref{Correlation coefficient wtd}. Unfortunately, the method presented in Eq.\eqref{FPTD Laplace Final} does not easily transfer to $F(n|\tau;n'|\tau_{n'})$: the conditional probability density that, given an initial tunneling to the drain, the jump number first reaches $n$ after a time $\tau_{n}$ and then first reaches $n'$ after another time $\tau_{n'}$. Ptaszynski has shown, however, that we can obtain the Pearson correlation coefficient from FPTDs of higher $n$ \cite{Ptaszynski2018}. The variance of $F(2|\tau)$ is 

\begin{align}
\langle \langle \tau_{2}^{2} \rangle \rangle^{*} &  = \langle \tau_{2}^{2}\rangle^{*} - (\langle \tau_{2} \rangle^{*})^{2}.
\end{align}
The average $\langle \hdots \rangle$ here implies an integral over all possible $\tau_{2}$; so that the second term, for example, is 
\begin{align}
(\langle \tau_{2} \rangle^{*})^{2} = \left( \int^{\infty}_{0} \: d\tau_{2} \: \tau_{2} F(2|\tau_{2}) \right)^{2}. \label{Pearson FPTD derivation mean squared}
\end{align}

We are interested in the joint distribution $F(1|\tau_{1} ; 2|\tau_{2})$, but we are $\text{\it{not}}$ searching for correlations between $\tau_{1}$ and $\tau_{2}$; we expect that $\tau_{2}$ will automatically be linearly correlated with $\tau_{1}$, since $\tau_{2} = \tau_{1} + \tau_{1'}$. Rather, we are searching for the correlation between $\tau_{1}$ and $\tau_{1'}$. With this in mind, we write 

\begin{align}
F(1|\tau{1} ; 2|\tau_{2}) & = F(1|\tau{1} ; 2|\tau_{1} + \tau_{1'})
\end{align}
and 
\begin{align}
F(1|\tau_{1}) & = \int_{0}^{\infty} \: d\tau_{1'} \: F(1|\tau_{1} ; 2|\tau_{1}+\tau_{1'}), \\ 
F(1|\tau_{1'}) & = \int_{0}^{\infty} \: d\tau_{1} \: F(1|\tau_{1} ; 2|\tau_{1}+\tau_{1'}), 
\end{align}
where the $F(1|\tau_{1'}$ is the probability density that the jump number first increases from $+1$ to $+2$ after a time delay of $\tau_{1'}$. From here we use the probabilistically self-evident identity, defined for $k = \{1,2,\hdots\}$, that 
\begin{align}
\langle \tau_{2}^k \rangle^{*} & =  \int^{\infty}_{0} \int^{\infty}_{0} d\tau_{1} \: d\tau_{1'}  \: (\tau_{1}+\tau_{1'})^k \: F(1|\tau_{1} ; 2|\tau_{1}+\tau_{1'});
\end{align}
and we obtain the relations 
\begin{align}
(\langle \tau_{2} \rangle^{*})^{2} & = (\langle \tau_{1} \rangle^{*})^{2} + 2\langle\tau_{1}\tau_{1'}\rangle^{*} + \langle \tau_{1'}\rangle^{*} \\
& = 2(\langle \tau_{1} \rangle^{*})^{2} + 2\langle\tau_{1}\tau_{1'}\rangle^{*}, \: \text{ and} \\
\langle \tau_{2}^{2} \rangle^{*} & = \langle \tau_{1}^{2}\rangle^{*} + 2\langle\tau_{1}\tau_{1'}\rangle^{*} + \langle \tau_{1'}^{2}\rangle^{*} \\
& = 2\langle \tau_{1}^{2}\rangle^{*} + 2\langle\tau_{1}\tau_{1'}\rangle^{*}, 
\end{align}
where the correlation function of two first-passage times is defined as
\begin{equation}
\langle\tau_{1}\tau_{1'}\rangle^{*} =\int^{\infty}_{0} \int^{\infty}_{0} d\tau_{1}d\tau_{1'} \: (\tau_{1}\tau_{1'})F(1|\tau_{1} ; 2|\tau_{1}+\tau_{1'}).
\end{equation}
Using the above averages, the Pearson coefficient is
\begin{align}
p^{*} & = \frac{\langle\tau_{1}\tau_{1'}\rangle^{*} - \langle\tau_{1}\rangle^{*}\langle\tau_{1'}\rangle^{*}}{\langle\langle \tau_{1}^{2} \rangle\rangle^{*}} \\
& = \frac{\langle\langle \tau_{2}^{2} \rangle\rangle^{*}}{2\langle\langle \tau_{1}^{2} \rangle\rangle^{*}} - 1, 
\end{align}

which is the correlation between when the jump number first reaches $+1$ and when it first reaches $+2$. 

We have seen that for equilibrated phonons the FPTD renewal assumption is satisfied, which means $F(1,2|\tau_{1},\tau_{2}) = F(1|\tau_{1}) \: F(1|\tau_{1})$, from Eq.\eqref{FPTD renewal 2}. This simplifies much of Eq.\eqref{Pearson FPTD derivation mean squared} and one can easily show that, as a result, $\langle\tau_{1}\tau_{1'}\rangle^{*} = \langle\tau_{1}\rangle^{*}\langle\tau_{1'}\rangle^{*}$. Evaluating the joint WTD in Eq.\eqref{second order WTD} for equilibrated phonons yields $\langle\tau\tau'\rangle = \langle\tau\rangle^{2}$ and clearly for both time distributions the Pearson correlation coefficient is $p = p^{*} = 0$.

\section{Connections} \label{Renewal and non-renewal Section}

From the previous sections, it is clear that there are a multitude of quantum statistics available, all describing the same transport scenario. An obvious question is whether all statistics provide complementary information, or whether, as is more interesting, there is information unique to each? We expect that, if fixed-time statistics contained identical information to fluctuating-time statistics, one should be able to reproduce the current cumulants from cumulants of the WTD or FPTD. 

For example, from physical intuition, the total average current is related to the average waiting time of the forward and backward distributions via 
\begin{align}
\langle I \rangle_{T} & = \langle I \rangle_{F} - \langle I \rangle_{B} \\
& = \frac{1}{\langle\tau\rangle_{F}} - \frac{1}{\langle\tau\rangle_{B}}, \label{average current from average waiting time}
\end{align}
where $\langle I \rangle_{F}$ and $\langle I \rangle_{B}$ are the forward and backward currents, respectively. Here, we see a relationship between the first cumulant of the directional current distribution and the first cumulant of the directional WTD. Two questions arise; firstly, is Eq.\eqref{average current from average waiting time} always true and if not under what conditions is it true; and secondly, can similar relations be found between all higher-order cumulants?

{{} These queries are neatly encapsulated by renewal theory. One can show that there exists exact relations between the FCS and cumulants of the WTD when the renewal assumption, Eq.\eqref{wtd-renewal}, is satisfied. Although Brandes initially demonstrated this for just a single-reset open quantum system  \cite{Brandes2008}, Budini \cite{Budini2011} and Albert et al. \cite{Albert2012} have shown that, under the renewal and unidirectional assumptions, the same one-to-one relations between WTD and FCS exist for multiple-reset systems. In the next section, we turn to the details of this derivation, following the works in Ref.[\onlinecite{Brandes2008}], Ref.[\onlinecite{Budini2011}], and Ref.[\onlinecite{Albert2011}]. The calculations are performed in the forward tunneling direction, but all results are equivalent for backward tunneling as well.

We start with the moment generating function of the current distribution:

\begin{align}
M(\chi,t) & = \sum_{n=0} e^{i n \chi}P(n,t), \label{MGF}
\end{align}
where the sum is for $n \geq 0$ since the transport is unidirectional. For $n > 0$ the probability $P(n,t)$ can be written generally in terms of the WTD:
\begin{widetext}
\begin{align}
P(n,t) & = \int_{0}^{t} \int_{0}^{t_{n-1}} \hdots \int_{0}^{t_{0}} \: d t_{n-1} d t_{n-2} \hdots d t_{0} \: w_{n}(t_{0},t_{1}-t_{0},\hdots,t_{n-1}-t_{n-2},t-t_{n-1})P(0,t_{0}). \label{General P(n,t) WTD}
\end{align}
\end{widetext}
In the stationary state the joint WTD does not depend on the initial time $t_{0}$. If the renewal assumption is satisfied, furthermore, then the joint WTD also factorises:
\begin{widetext}
\begin{align}
P(n,t) & = \int_{0}^{t} \int_{0}^{t_{n-1}} \hdots \int_{0}^{t_{0}} \: d t_{n-1} d t_{n-2} \hdots d t_{0} \: w(t-t_{n-1})w(t_{n-1}-t_{n-2}) \hdots w(t_{1}-t_{0}) P(0,t_{0})
\end{align}
\end{widetext}
Recognising that $P(1,t) = \int_{0}^{t} \: d t_{0} w(t-t_{0})P(0,t_{0})$, and so on, $P(n,t)$ can now be written recursively as 

\begin{align}
P(n,t) & = \int_{0}^{t} \: d t_{n-1} \: w(t-t_{n-1}) P(n-1,t_{n-1}). \label{renewal recursive equation}
\end{align}

Eq.\eqref{renewal recursive equation} is now inserted into the moment generating function to obtain
\begin{widetext}
\begin{align}
M(\chi,t) & = P(0,t) + \sum_{n=1}^{\infty} \int_{0}^{t} \: d t_{n-1} \: w(t-t_{n-1}) e^{i n \chi} P(n-1,t_{n-1}) \\
& = P(0,t) + e^{i\chi} \int_{0}^{t} \: d t_{n-1} \: w(t-t_{n-1}) M(\chi,t_{n-1}).
\end{align}
\end{widetext}
As usual with convolution integrals, it is easier to work in Laplace space:
\begin{align}
\tilde{M}(\chi,z) & = \tilde{P}(0,z) + e^{i\chi} \tilde{w}(z) \tilde{M}(\chi,z),
\end{align}
Rearranging gives us
\begin{align}
\tilde{M}(\chi,z) & = \frac{\tilde{P}(0,z)}{1 - e^{i\chi} \tilde{w}(z)},
\end{align}
which in time-space is given by the inverse Laplace transform
\begin{align}
M(\chi,t) & = \frac{1}{2\pi i}\int_{c-i\infty}^{c+i\infty} \tilde{M}(\chi,z) =\sum_{z_{k}} \text{Res}\left(\tilde{M}(\chi,z),z_{k}\right). 
\label{General residue equation}
\end{align}
Examining $\tilde{M}(\chi,z)$, we see that the poles are those values $\left\{z_{k}(\chi)\right\}$ that satisfy the equation 

\begin{align}
0 & = 1 - w(z_{k}(\chi))e^{i\chi}. \label{Equation to be satisfied WTD}
\end{align}
If this equation has one solution, corresponding to a simple pole $z_{0}(\chi)$, then the integral is easily evaluated:
\begin{align}
M(\chi,t) & = \lim_{z\rightarrow z_{0}(\chi)} (z - z_{0}(\chi)) \frac{\tilde{P}(0,z)e^{zt}}{1 - e^{i\chi} \tilde{w}(z)} \\
& = \tilde{P}(0,z_{0}(\chi))e^{z_{0}(\chi)t}. \label{before long-time limit}
\end{align}
We now write the moment generating function in terms of the cumulant generating function, as in Eq.\eqref{MGF from CGF}, and note that in the long-time limit, as $t\rightarrow \infty$, the cumulant generating function is given by Eq.\eqref{eigenvalue simplification}, the large-deviation principle:
\begin{align}
\lim_{t\rightarrow\infty} M(\chi,t) & = \lim_{t\rightarrow\infty} e^{K(\chi,t)} \\ 
& \approx e^{\Lambda_{max}t}.
\end{align}

Applying the long-time limit to Eq.\eqref{before long-time limit} as well, we see that the exponential term dominates $\tilde{P}(0,z_{0}(\chi))$ and so in this limit $\Lambda_{max} = z_{0}(\chi)$.

We now demonstrate  that the same result also holds in the case when $\tilde{M}(\chi,z)$ poses multiple poles of any order.  Suppose that $\tilde{M}(\chi,z)$ has $M$ poles $z_0(\chi), \hdots, z_{M-1} (\chi)$, where $z_0(\chi) $ is the the dominant pole with the largest real part. For a  pole $z_{k}(\chi)$  of order $m$, the residue in Eq.\eqref{General residue equation} is 
\begin{align}
 & \frac{1}{(m-1)!} \frac{d^{m-1}}{dz^{m-1}}\left[(z-z_{k}(\chi))\tilde{M}(\chi,z)\right]\Bigg|_{z=z_{k}(\chi)} \\
& =  \alpha_{z_{k}(\chi)}(t) e^{t z_{k}(\chi)},
\end{align}
 where $\alpha_{z_{k}(\chi)}(t)$, a polynomial of order $m-1$,  comes from evaluating the successive $(m-1)$ derivatives at $z=z_{k}(\chi)$.
The moment generating function is then given by Eq.\eqref{General residue equation}, which  is 
\begin{multline}
M(\chi,t)  = \sum^{M-1}_{k=0} \alpha_{z_{k}(\chi)}(t) e^{t z_{k}(\chi)} 
\\
= \alpha_{z_{0}(\chi)}(t) e^{t z_{0}(\chi)} \Big[ 1+ \sum^{M-1}_{k=1} \alpha_{z_{k}(\chi)}(t) e^{t z_{k}(\chi)/z_{0}(\chi)} \Big].
\end{multline}
In the long-time limit the exponential term $e^{t z_{0}(\chi)}$ dominates the expression, yielding 
\begin{equation}
M(\chi,t)  \approx  e^{t z_{0}(\chi)}
\end{equation}
Therefore,  regardless of the nature of the poles, in the long-time limit we are left with a dominant solution of Eq.\eqref{Equation to be satisfied WTD}: $z_{0}(\chi) = \Lambda_{max}(\chi)$. 

With this information, we rewrite Eq.\eqref{Equation to be satisfied WTD} in the long-time limit as
\begin{align}
0 & = i\chi + \ln \tilde{w}(z)\Bigg|_{z=\Lambda_{max}(\chi)}. \label{Relationship equation}
\end{align}
We note that at $\chi=0$, $\mathbf{L}(\chi) = \mathbf{L}$ and $\Lambda_{max}(0)=0$, since all other eigenvalues are negative due to the structure of the Liouvillian. Let us now take successive derivatives of Eq.\eqref{Relationship equation} with respect to $\chi$ and then set $\chi=0$, as we do when generating current cumulants in Eq.\eqref{full current cumulant definition}. The first derivative yields

\begin{align}
0 & = -i\frac{\partial}{\partial\chi}\left[i\chi + \ln \tilde{w}(z)\right]\Bigg|_{z=\Lambda_{max}(\chi); {\chi=0}} \\
& = 1 - i \frac{\partial z}{\partial \chi}\frac{\partial \ln\tilde{w}}{\partial z}\Bigg|_{z=\Lambda_{max}(\chi); {\chi=0}}  \\
& = 1 - i \frac{\partial \Lambda_{max}}{\partial \chi}\Bigg|_{\chi=0}\:\frac{\partial \ln\tilde{w}}{\partial z}\Bigg|_{z=0}. \label{First derivative relationship working}
\end{align}

Comparing Eq.\eqref{First derivative relationship working} with the definitions of the WTD cumulants and the FCS in Eq.\eqref{WTD cumulants} and Eq.\eqref{full current cumulant definition}, respectively, we see that it contains the first cumulants of both distributions:

\begin{align}
0 & = 1 - \langle\langle I \rangle\rangle \: \langle\langle \tau \rangle\rangle, & \text{ so that } \\
\langle\langle I \rangle\rangle & = \frac{1}{\langle\langle \tau \rangle\rangle}, \label{first cumulant equivalence}
\end{align}
which is the intuitive relationship outlined earlier in Eq.\eqref{average current from average waiting time}. Taking the second derivative we get 

\begin{align}
0 & = (-i)^{2}\frac{\partial^{2}}{\partial\chi^{2}}\left[i\chi + \ln \tilde{w}(z)\right]\Bigg|_{z=\Lambda_{max}(\chi); {\chi=0}}  \\
0 & = \frac{\partial}{\partial\chi}\left[\frac{\partial z}{\partial \chi}\frac{\partial \ln\tilde{w}}{\partial z}\right]\Bigg|_{z=\Lambda_{max}(\chi); {\chi=0}}  \\
0 & = \frac{\partial^{2} \Lambda_{max}}{\partial \chi^{2}}\Bigg|_{\chi=0}\frac{\partial \ln\tilde{w}}{\partial z}\Bigg|_{z=0} + \left(\frac{\partial \Lambda_{max}}{\partial \chi}\right)^{2}\Bigg|_{\chi=0}\frac{\partial^{2} \ln\tilde{w}}{\partial z^{2}}\Bigg|_{z=0},
\end{align}
which, after comparison with the definitions of the second-order current and waiting time cumulants, reduces to 

\begin{align}
\frac{\langle\langle I^{2} \rangle\rangle}{\langle\langle I \rangle\rangle} & = \frac{\langle\langle \tau^{2} \rangle\rangle}{\langle\langle \tau \rangle\rangle^{2}}. \label{second cumulant equivalence}
\end{align}
Continuing, we get all relationships between higher-order cumulants as well; the skewness, for example, is

\begin{align}
\frac{\langle\langle I^{3} \rangle\rangle}{\langle\langle I \rangle\rangle} & = 3\frac{\langle\langle \tau^{2} \rangle\rangle^{2}}{\langle\langle \tau \rangle\rangle^{4}} - \frac{\langle\langle \tau^{3} \rangle\rangle}{\langle\langle \tau \rangle\rangle^{3}}, \label{third cumulant equivalence}
\end{align}
and so on. The LHS of Eq.\eqref{second cumulant equivalence} is the Fano factor and the RHS is the randomness parameter. These quantities, therefore, provide a direct test of whether the transport is renewal; we can plot the Fano factor alongside the randomness parameter and identify non-renewal behaviour where they deviate.} 

This mapping between the two sets of statistics unfortunately does not hold for bidirectional transport. Furthermore, although we can define relations between the cumulants of the WTD and current distribution in either the forward or backward direction, we cannot combine them to reproduce the appropriate cumulants of the total current distribution $\langle\langle I^{k}\rangle\rangle_{T}$. The obvious exception is the physically evident relation for the average current in Eq.\eqref{average current from average waiting time}. Not only does it relate the first cumulant of the directional WTDs to the first cumulant of the total current distribution, but it has the additional property that it is true regardless of whether the renewal assumption is satisfied. To see, we will consider the forward current $\langle I \rangle_{F}$, which is reconstructed from the WTD by assuming that the set of available currents is $\left\{ \frac{k}{\langle\tau_{1}+ \hdots + \tau_{k}\rangle_{F}}\right\}$, each occuring with probability $P(k)$. The average forward current is then 

\begin{align}
\langle I \rangle_{F} & = \sum_{m=1}^{\infty} \frac{k}{\langle\tau_{1}+ \hdots + \tau_{k}\rangle_{F}} P(k), \label{average current reconstructed from WTD}
\end{align}
and the average waiting times can be simplified without using the renewal assumption:
\begin{widetext}
\begin{align}
\langle \tau_{1} + \tau_{2} + \hdots + \tau_{k}\rangle_{F} & = \int_{0}^{\infty} d\tau_{k} \hdots \int_{0}^{\infty} d\tau_{2}\int_{0}^{\infty} d\tau_{1} \: (\tau_{1} + \tau_{2} + \hdots + \tau_{k}) \: w_{F}(\tau_{1},\tau_{2}\hdots,\tau_{k}). \\
& = \sum_{m=1}^{k} \int_{0}^{\infty} d\tau_{m} \: \tau_{m} \: w_{F}(\tau_{m}) \\
& = k\langle \tau \rangle_{F}.
\end{align}
\end{widetext}
Eq.\eqref{average current reconstructed from WTD} then reduces to 
\begin{align}
\langle I \rangle_{F} & = \frac{1}{\langle \tau \rangle_{F}},
\end{align}
since $\sum\limits_{k=1}^{\infty} P(k) = 1$. The backward current can be similarly defined, and thus Eq.\eqref{average current from average waiting time} is satisfied. We cannot do the same for higher-order cumulants, since for $k>1$ we cannot write $\langle\langle I^{k} \rangle\rangle$ as an explicit reconstruction from the direction WTDs. In renewal theory, then, the WTD is evidently limited to unidirectional transport.

It has recently been shown, however, that when the renewal assumption is satisfied, similar relations can be found between the FCS and cumulants of the FPTD that hold even when the transport is bidirectional \cite{Ptaszynski2018}. We will not reproduce the derivation here, but rather direct the reader to Ref.[\onlinecite{Ptaszynski2018}] for an explicit overview. In Eq.\eqref{FPTD cumulant relations} we saw that for renewal transport all FPTD cumulants are linearly related, so the relations between current cumulants and FPTD cumulants can all be expressed using $F(1|\tau)$:

\begin{gather}
\langle I \rangle_{T} = \frac{n}{\langle\tau_{n}\rangle^{*}} =\frac{1}{\langle\tau_{1}\rangle^{*}} \label{Average currrent from FPTD cumulants} \\
\frac{\langle\langle I^{2} \rangle\rangle_{T}}{\langle I \rangle_{T}} = n\frac{\langle\langle\tau_{n}^{2}\rangle\rangle^{*}}{(\langle\tau_{n}\rangle^{*})^{2}} = \frac{\langle\langle \tau_{1}^{2}\rangle\rangle^{*}}{(\langle\tau_{1}\rangle^{*})^{2}}, \label{Fano from FPTD}
\end{gather}
which hold even when the transport is bidirectional. Eq.\eqref{Fano from FPTD} implies that, if the renewal assumption is satisfied, then $|F - R^{*}| = 0$. Examining the exact results in Eq.\eqref{FPTD exact RP} and Eq.\eqref{FCS exact Fano}, we see that the Fano factor and FPTD randomness parameter do indeed match.

We might expect that, since Eq.\eqref{Average currrent from FPTD cumulants} is analogous to Eq.\eqref{average current reconstructed from WTD}, it also holds regardless of the renewal assumption. In this case, though, the average current is reconstructed from the FPTD as 

\begin{align}
\langle I \rangle_{T} & = \sum_{k=1}^{\infty} \frac{k}{\langle\tau_{k}\rangle^{*}} P(k),
\end{align}
and this cannot be simplified since for non-renewal statistics $\langle\tau_{k}\rangle \neq k\langle\tau_{1}\rangle$ and in general $F(k|\tau) \neq F(1|\tau_{1} ; \hdots ; 1|\tau_{1}^{(k)})$, unless the transport is unidirectional.

As we saw in Section (\ref{Higher-order time distributions}) non-renewal transport is accompanied by temporal correlations in the first-passage and waiting times. This is an example of the unique information only available to the fluctuating-time distributions when the renewal assumption is violated.

\section{Illustration on the Holstein model} \label{Example Section}

Throughout the paper we have demonstrated that, for phonons in equilibrium, the FCS, WTD, and FPTD cumulants can all be analytically derived. Furthermore, the exact relation between the FPTD cumulants and the FCS demonstrated that transport in this regime satisfies the renewal assumption and subsequent waiting and first-passage times are uncorrelated: $p = p^{*} = 0$. While the equilibrated phonon example provides an analytic demonstration, it does not provide a qualitatively interesting example since all correlations are zero. When the phonons are unequilibrated, however, the transport picture is more complicated. In this section we use quantitative results to demonstrate how one can use fluctuation statistics to obtain information about quantum transport, in particular the renewal behavior.

\begin{figure*}
	\subfloat[]{\includegraphics[scale=0.5]{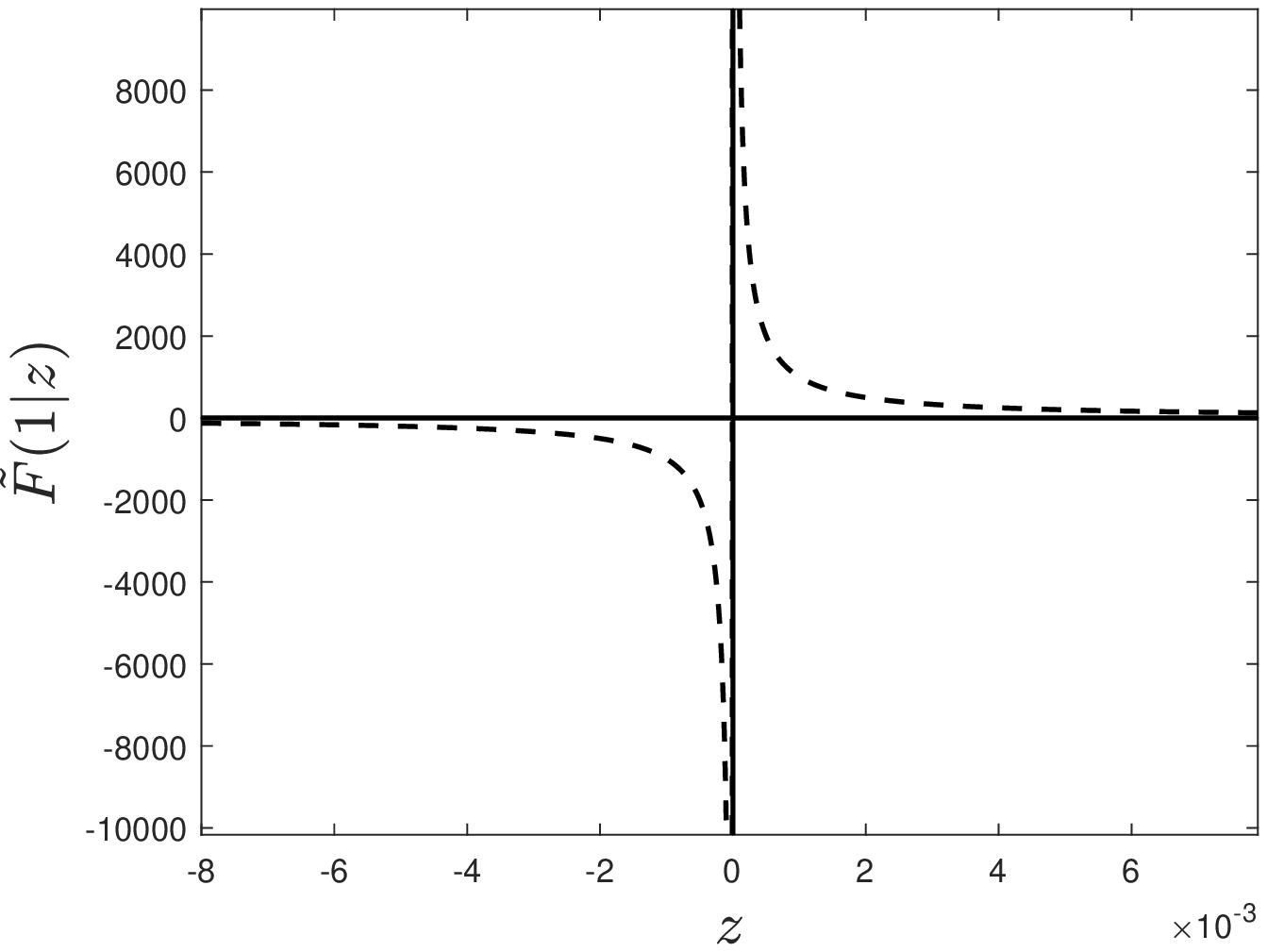}\label{Fig_FPTD_Laplace_a}}
	\subfloat[]{\includegraphics[scale=0.5]{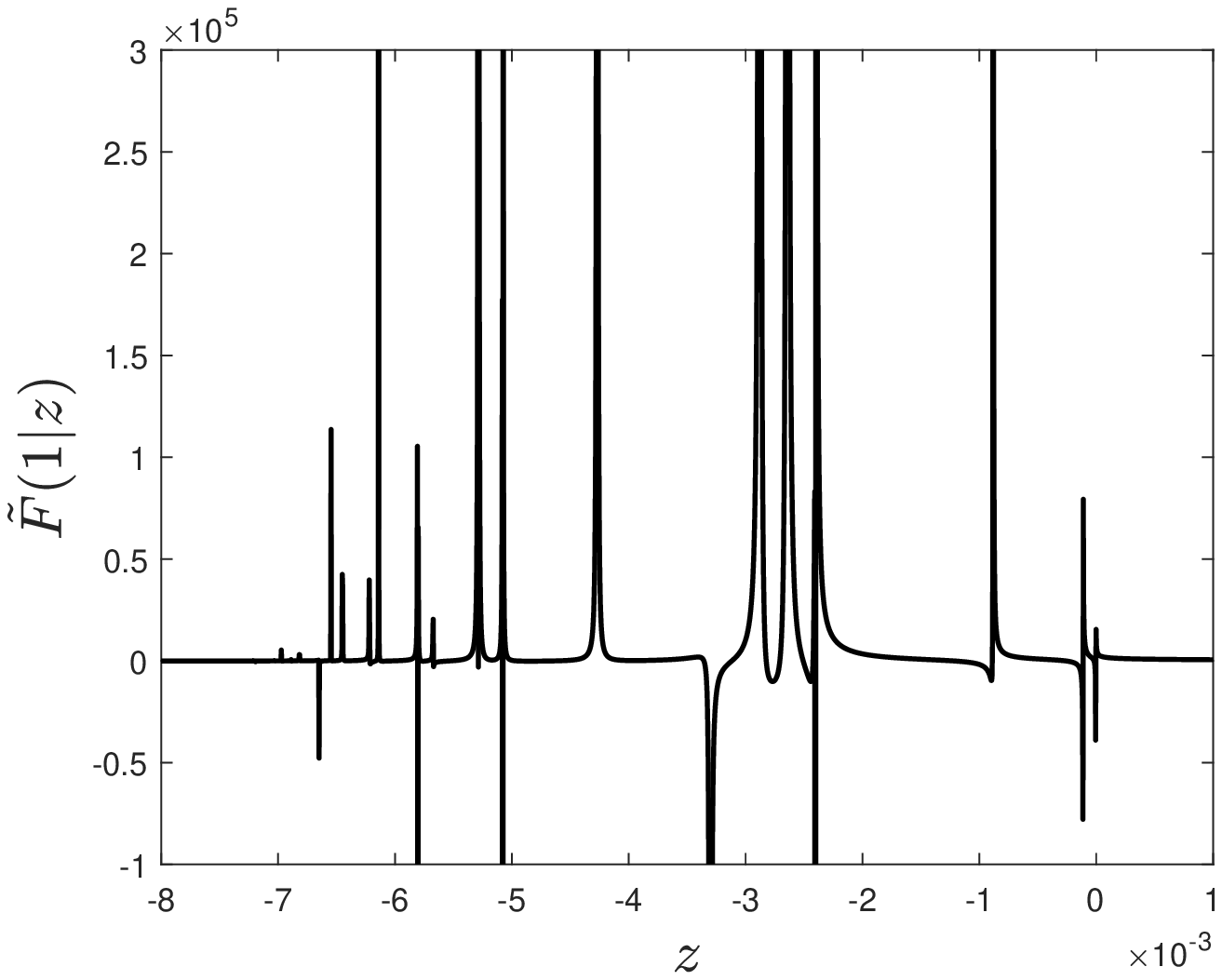}\label{Fig_FPTD_Laplace_b}}
	\caption{The FPTD for transport through the Holstein model, where (a) the phonons are in equilibrium and (b) the phonons are out-of-equilibrium. The vibrationally adjusted energy level is $\varepsilon = 0$, the electron-phonon coupling is $\lambda=4$, the vibrational frequency is $\omega = 1$, $T = 0.05$, the equilibrium phonons are kept at vibrational temperature $T_{V}=0.05$, and $\gamma_{\alpha} = \frac{\gamma}{2} = 0.01$. We use units of $\omega$ for all energy parameters (or $\hbar\omega/e$  if we reintroduce $\hbar\text{ and }e$).}
\end{figure*}

In Fig.(\ref{Fig_FPTD_Laplace_a}) and Fig.(\ref{Fig_FPTD_Laplace_b}), in which the FPTD in Laplace space is plotted for equilibrated and unequilibrated phonons, we see direct evidence of the complexity difference. While in general the first-passage time in time space is difficult to compute, due to the demanding numerical Bromwich integral, one can obtain information from the FPTD in Laplace space. In  Fig.(\ref{Fig_FPTD_Laplace_a}) and Fig.(\ref{Fig_FPTD_Laplace_b}), the chemical potentials are $\mu_{S} = 4.5\text{ and }\mu_{D} = -4.5$. At this voltage, backscattering effects are minimised as $\mu_{D}$ is off-resonance with the vibrationally shifted energy levels and $\tilde{F}(1|z)$ will resemble $\tilde{w}_{F}(z)$. For equilibrated phonons, the definition in Eq.\eqref{Forward WTD definition Laplace} indicates that the FPTD behavior will be determined by two poles at 

\begin{align}
z = -(T_{10}+T_{01}) \pm \sqrt{(T_{10}+T_{01})^{2} - 4T_{10}T_{01}^{D}};
\end{align}
however, for the parameters chosen, $T_{01} \approx T_{01}^{D}$ and the transport is dominated by one pole at approximately $-(T_{10}+T_{01})$. This is the behavior observed in Fig.(\ref{Fig_FPTD_Laplace_a}). The FPTD for unequilibrated phonons, shown in Fig.(\ref{Fig_FPTD_Laplace_b}), is by contrast much more complicated with multiple poles and additional side peaks. To obtain more direct information we must turn to cumulant analysis.

\begin{figure*}
	\subfloat[]{\includegraphics[scale=0.5]{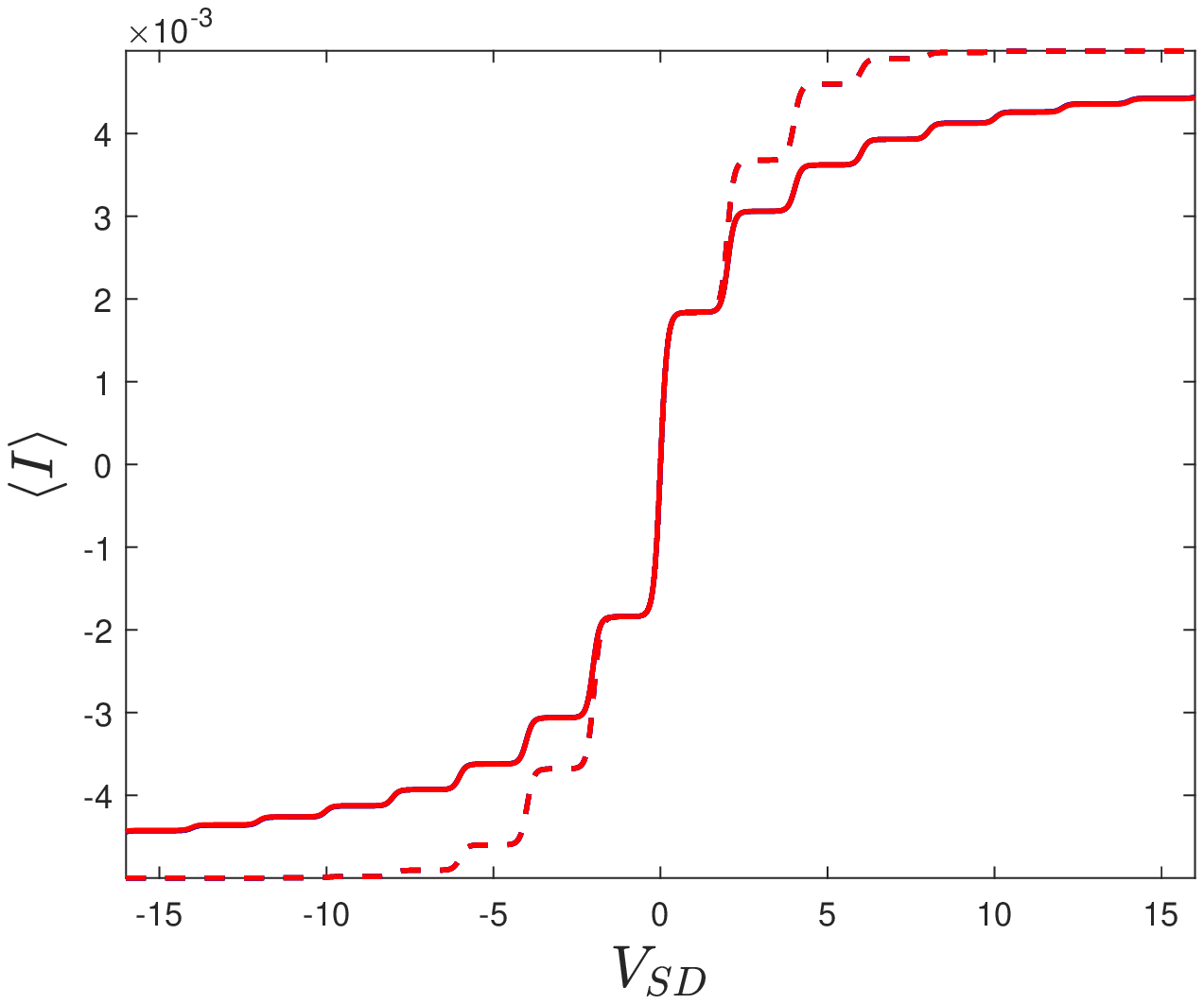}\label{Fig_Current_a}}
	\subfloat[]{\includegraphics[scale=0.5]{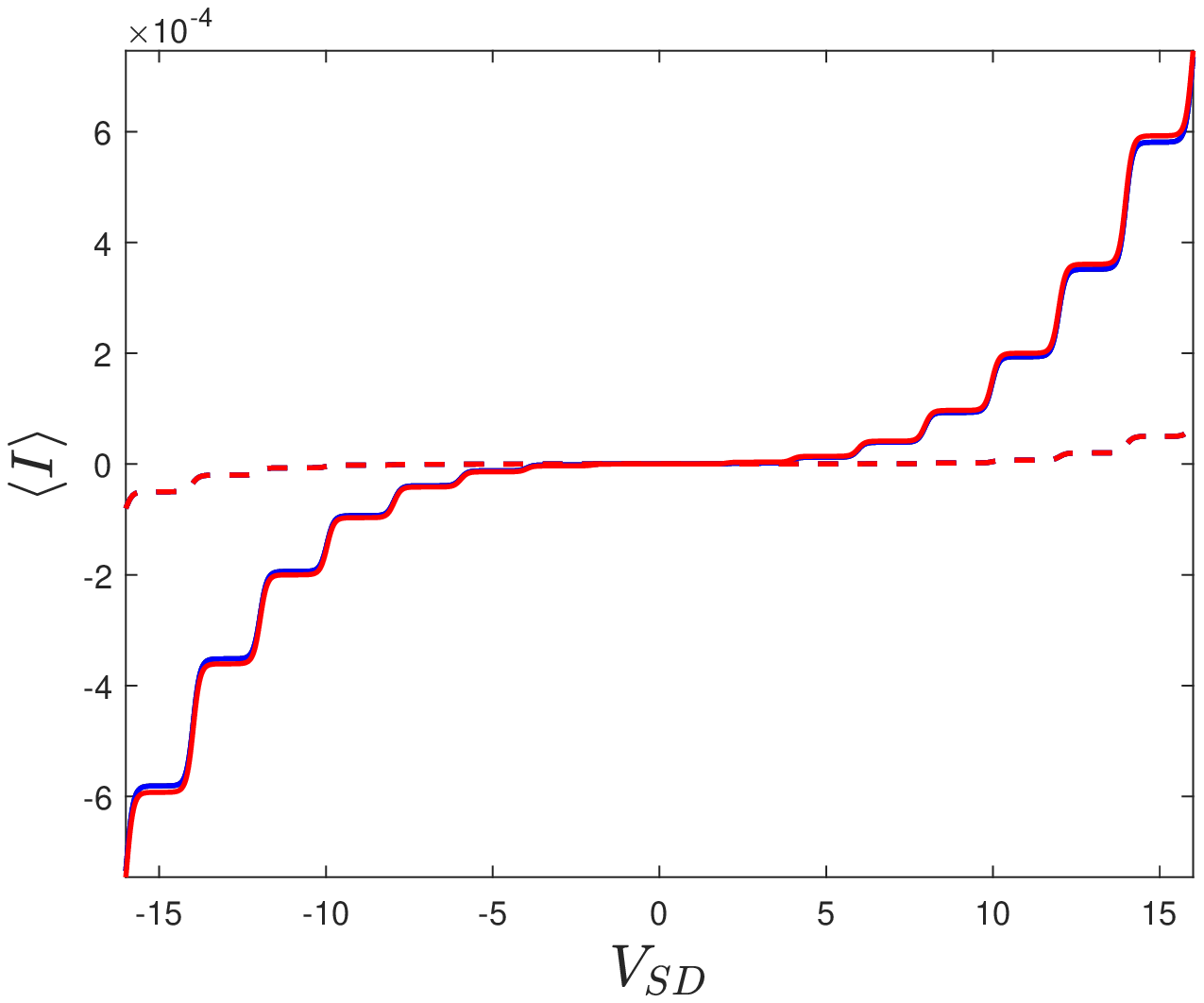}\label{Fig_Current_b}}
\caption{Color online. Exact current for equilibrium and non-equilibrium phonons $\langle I \rangle = \left(\mathbf{I},(\mathbf{J}_{F} - \mathbf{J}_{B})\bar{\mathbf{P}}\right)$ (black) compared to the respective predictions from the FPTD $\frac{1}{\langle\tau\rangle^{*}}$ (red) and WTD $\frac{1}{\langle\tau\rangle_{F}} - \frac{1}{\langle\tau\rangle_{B}}$ (blue). In (a) the 
electron-phonon interaction is $\lambda = 1$, while in (b) $\lambda = 4$. For $V_{SD} < 0$ the FPTD current is $-\frac{1}{\langle \tau_{(-1)} \rangle^{*}}$ and for $V_{SD} > 0$ it is $\frac{1}{\langle \tau_{1} \rangle^{*}}$. The vibrationally adjusted energy level is $\varepsilon = 0$, the vibrational frequency is $\omega = 1$, $T = 0.05$, $\gamma_{\alpha} = \frac{\gamma}{2} = 0.01$, and the equilibrium phonons are kept at a vibrational temperature of $T_{V} = 0.05$. The source and drain chemical potentials are shifted symmetrically about zero: $\mu_{S} = -\mu_{D} = V_{SD}/2$. We use units of $\omega$ for $\langle I \rangle$ (or $e\omega$) and units of $\omega$ for all energy parameters.}
\end{figure*}

{{} Fig.(\ref{Fig_Current_a}) and Fig.(\ref{Fig_Current_b}) are reconstructions of well-known current phenomena; see, for example, Ref.[\onlinecite{Mitra2004}] and Ref.[\onlinecite{Flindt2005}]. Alongside the current, our plots include predictions of the current from the WTD, $\frac{1}{\langle\tau\rangle_{F}} - \frac{1}{\langle\tau\rangle_{F}}$, and the FPTD $\frac{1}{\langle\tau_{1}\rangle^{*}}$.} For a strong electron-phonon coupling $\lambda \gg 1$, shown in Fig.(\ref{Fig_Current_b}), the Franck-Condon blockade suppresses the current at low voltages \cite{Koch2005}, as in this regime $|X_{qq'}|$ is minimised for small $|q-q'|$. Evidently current is suppressed more for equilibrated phonons than unequilibrated, as the high energy transitions needed to overcome the blockade are more likely when the phonons are out-of-equilibrium. For a moderate electron-phonon coupling $\lambda \sim 1$, shown in Fig.(\ref{Fig_Current_a}), the blockade disappears and forcing the phonons to equilibrium actually increases the current; transport is dominated by elastic or small $|q-q'|$ transitions, which are more populated in equilibrium \cite{Koch2005}. To draw connections with fluctuating-time statistics, we have superimposed plots of the total current reconstructed from the WTD, outlined in Eq.\eqref{average current from average waiting time}, and from the FPTD, outlined in Eq.\eqref{Average currrent from FPTD cumulants}.

\begin{figure*}
{\includegraphics[scale=0.5]{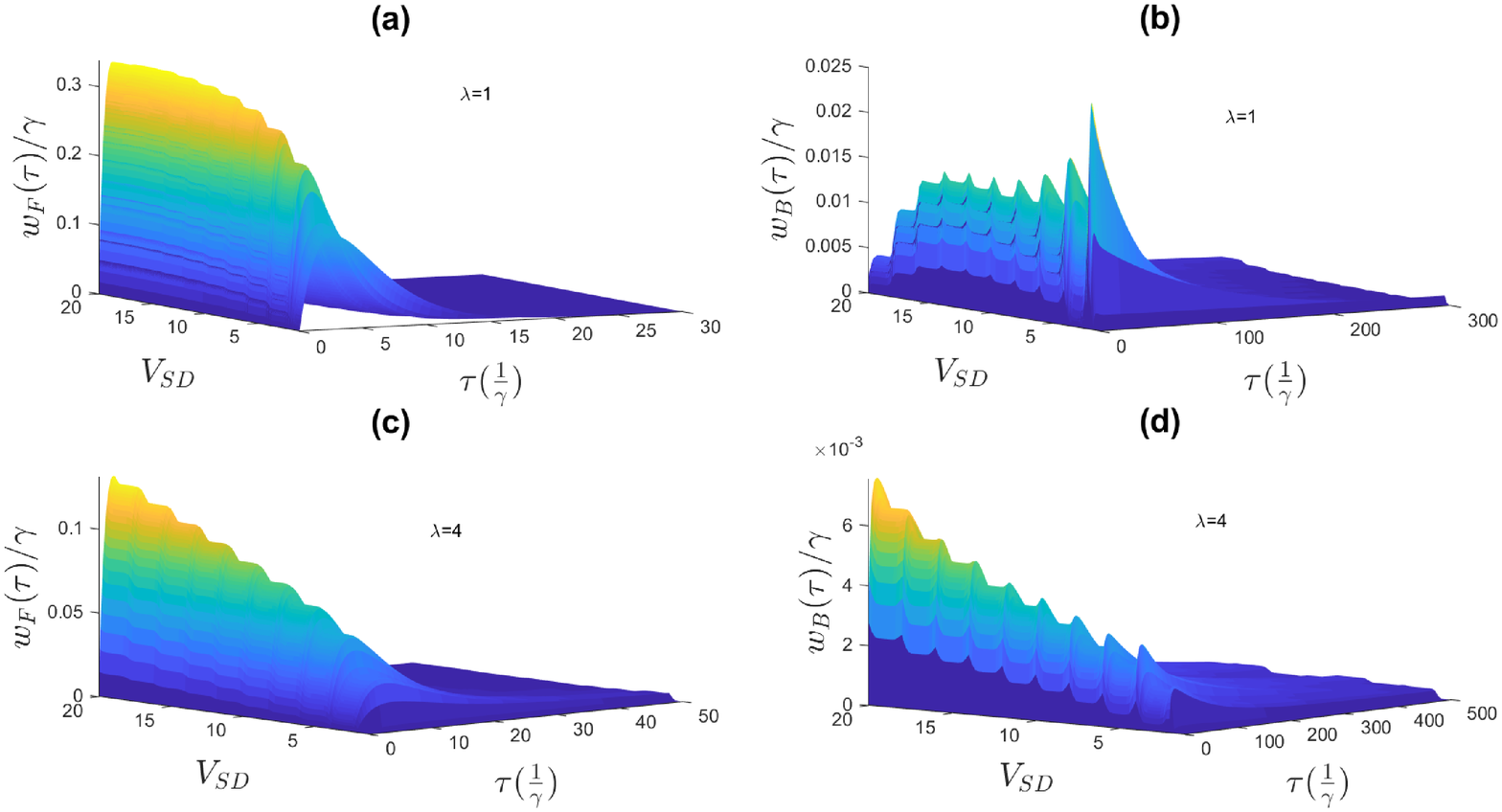}} 
\caption{Color online. Contour plots of the WTD as a function of voltage and time, for transport through the Holstein model for unequilibrated phonons and for different $\lambda$. In (a) and (c) $w_{F}(\tau)$ is plotted for $\lambda = 1$ and $\lambda = 4$, respectively. Likewise, in (b) and (d) $w_{B}(\tau)$ is plotted for $\lambda = 1$ and $\lambda = 4$, respectively. The WTD is calculated using Eq.\eqref{brandes definition} and the respective choices of jump operator: $\mathbf{J}_{F}$ and $\mathbf{J}_{B}$. The vibrationally adjusted energy level is $\varepsilon = 0$, the vibrational frequency is $\omega = 1$, $T = 0.05$, and $\gamma_{\alpha} = \frac{\gamma}{2} = 0.01$. The source and drain chemical potentials are shifted symmetrically about zero: $\mu_{S} = -\mu_{D} = V_{SD}/2$. We use units of $\omega$ for all energy parameters (or $\hbar\omega/e$).}
\label{Fig_WTD_NE.eps}
\end{figure*}

\begin{figure}
{\includegraphics[scale=0.5]{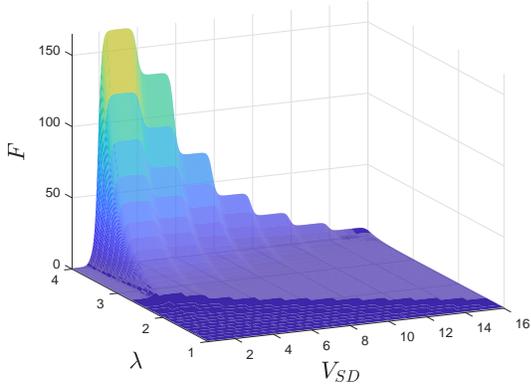}} 
\caption{Color online. Surface plot of the Fano factor, calculate using Eq.\eqref{FF n definition}, as a function of source-drain bias voltage $V_{SD}$ and electron-phonon coupling $\lambda$. The vibrationally adjusted energy level is $\varepsilon = 0$, the vibrational frequency is $\omega = 1$, $T = 0.05$, and $\gamma_{\alpha} = \frac{\gamma}{2} = 0.01$. The source and drain chemical potentials are shifted symmetrically about zero: $\mu_{S} = -\mu_{D} = V_{SD}/2$. We use units of $\omega$ for all energy parameters.}
\label{Fig_Fano_surface}
\end{figure}

\begin{figure*}
{\includegraphics[scale=0.5]{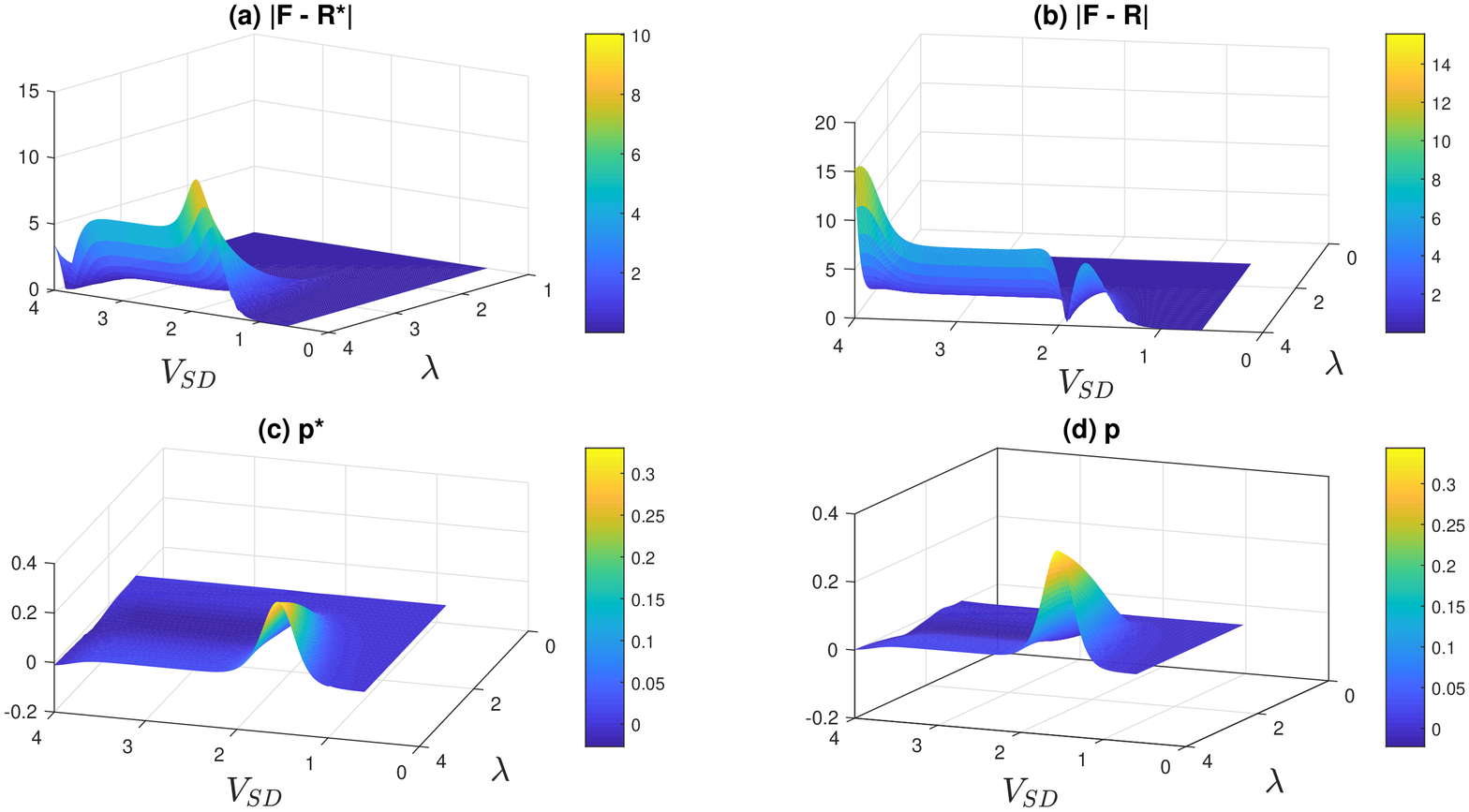}}
\caption{Color online. Color plots of (a) and (b) the absolute difference between the Fano factor and the WTD and FPTD randomness parameter, respectively, and (c) and (d) the corresponding Pearson correlation coefficients, all as functions of both $\lambda$ and voltage. All plots describe transport through the Holstein model, where phonons are unequilibrated. The WTD randomness parameter $R$ is calculated only from the forward WTD. The vibrationally adjusted energy level is $\varepsilon = 0$, the vibrational frequency is $\omega = 1$, $T = 0.05$, $\gamma_{\alpha} = \frac{\gamma}{2} = 0.01$. The source and drain chemical potentials are shifted symmetrically about zero: $\mu_{S} = -\mu_{D} = V_{SD}/2$. We use  units of $\omega$ for all energy parameters.}
\label{Fig_Fano_Pearson}
\end{figure*}
We saw that, regardless of whether the renewal assumption is satisfied, the total current is always exactly reproduced by the directional WTDs: $\langle I \rangle_{T} = \frac{1}{\langle\tau\rangle_{F}} - \frac{1}{\langle\tau\rangle_{B}}$. Both Fig.(\ref{Fig_Current_a})  and Fig.(\ref{Fig_Current_b}) confirm this as the blue WTD exactly matches the exact current, plotted in black. The FPTD, on the other hand, exactly reproduces the total current only when the renewal assumption is satisfied. At moderate $\lambda$ the renewal assumption appears satisfied for all voltages, which is physically plausible since the inelastic scatterings, a key source of non-renewal behavior, are minimised for $\lambda=1$. In Fig.(\ref{Fig_Current_b}), however, the FPTD reproduction differs from the actual current at large $|V_{SD}|$, indicating non-renewal transport in this regime.

The final interesting current feature is the characteristic steplike structure, where each step corresponds to the opening of another conduction channel. These occur when the voltage reaches multiples of $2\omega$: $V_{SD} = 2q\omega$, when high- or low-energy electrons may interact with $q$ vibrational quanta and either lose or gain the exact energy required to resonantly tunnel  through the $\varepsilon = 0$ level \cite{Mitra2004,scheer2010molecular,Koch2005}. This behavior is reflected in the WTDs, plotted in Fig.(\ref{Fig_WTD_NE.eps}). For simplicity we have plotted over the positive voltage range only, so that forward transitions are preferred by the transport. As the voltage increases, the mode of $w_{F}(\tau)$ and the mean time $\langle\tau\rangle_{F}$ decrease at multiples of $2\omega$, corresponding to current steps at these voltages. The Franck-Condon blockade is also visible, as for $\lambda = 1$ the WTD narrows about its peak for much smaller voltages than for $\lambda = 4$. 

The WTDs for backward tunnelings also display steplike changes, alongside peaks at certain voltages. When $\mu_{D}$ is in resonance with one of the quasi-energy levels $\varepsilon - q\omega$, an effective backtunneling conduction channel is opened. This channel is unavailable at other voltages due to the quantisation of the vibrational energy. For moderate couplings this effect disappears at high voltages, as the electron does not spend enough time in the molecule to interact with high-energy phonons. In contrast, for $\lambda = 4$ the effect increases as voltage increases but is minimised at low voltages due to the Franck-Condon blockade.

Identifying non-renewal behavior by visually inspecting the average current can be misleading, as small but important numerical differences can be invisible to the naked eye. Indeed, while it appears that, for small voltages and when $\lambda = 4$, the FPTD exactly reproduces $\langle I \rangle_{T}$ and thus the transport is renewal, there are actually significant correlations between successive waiting and first-passage times. These are masked by the small currents at play; to unveil we compare the Fano factor and randomness parameters, which are scaled by $\langle I \rangle$ and so all differences are detectable. 

The Fano factor, as a function of the $\lambda$ and $V_{SD}$, is plotted in Fig.(\ref{Fig_Fano_surface}). We have omitted the regime $V_{SD} < 0.3$ as, at low $V_{SD}$ and no temperature gradient, $F$ diverges for any molecular system, since $\langle I \rangle \rightarrow 0$. For many molecular systems the Fano factor is $\sim 1$; however, previously Koch and von Oppen \cite{Koch2005} have shown that unequilbrated phonon transitions and a strong electron-phonon coupling result in electron bunching, effectively widening the current distribution so that $\langle\langle n^{2} \rangle\rangle \gg \langle n \rangle$. Fig.(\ref{Fig_Fano_surface}) shows that as $\lambda$ decreases this effect vanishes alongside the Franck-Condon blockade. 

Between $V_{SD} = 1$ and $V_{SD} = 2$, and for strong $\lambda$, successive waiting and first-passage times are highly positively correlated, which is an entirely separate phenomenon to the electron bunching evident in the Fano factor. These have previously been explained by D.S.K with an elastic ``shortcut'' channel through the $q=3$ vibrational state, which opens when the first waiting time is small \cite{Kosov2017b}. We can see this non-renewal transport either from the difference between the Fano factor and the randomness parameters,  $|F-R^{*}|$ and $|F-R|$, or directly from the Pearson correlation coefficient: all are shown in Fig.(\ref{Fig_Fano_Pearson}).

From the discussion on renewal theory, we know that $|F - R^{*}| \neq 0$ and $|F - R| \neq 0$ only when the transport is non-renewal. The difference between $F$ and $R$, however, can only be used to identify non-renewal transport for unidirectional transitions. We can see from Fig.(\ref{Fig_Fano_Pearson}b) that for $\lambda = 4$ and just below $V_{SD} = 2$ the difference $|F-R|$ is zero, implying that the transport is renewal, but the correlations in Fig.(\ref{Fig_Fano_Pearson}d) are nonzero: a direct contradiction. In contrast, the difference $|F-R^{*}|$ in Fig.(\ref{Fig_Fano_Pearson}a) is nonzero when there are nonzero correlations in Fig.(\ref{Fig_Fano_Pearson}c); in particular, when $\lambda = 4$ and just below $V_{SD} = 4$ the difference $|F-R^{*}| = 0$ and $p^{*} = 0$. We note that for 2-dimensional plots the differences can sometimes be hard to see. Overall then, the first-passage time cumulants are a more general method for identifying non-renewal transport.

\section{Conclusions} \label{Conclusions}

In this paper we aimed to provide a pedagogical and methods based approach to fluctuations in mesoscopic quantum transport: summarising and explaining fixed-time and fluctuating-time statistics in a master equation framework. We focused on three main statistics: the FCS, the WTD, and the FPTD. All statistics were calculated via a Markovian master equation for the demonstrative example of transport through the Holstein model, in which a single electronic energy level interacts with vibrational phonons. We demonstrated how to calculate cumulants of the current distribution, the FCS, as well as cumulants of the WTD and FPTD; analytic results were available in the case of equilibrated phonons but numerical evaluation was required for the fully non-equilibrium case. We saw that when the renewal assumption is satisfied there exists direct relationships between FPTD cumulants and the FCS as well as between WTD cumulants and the FCS, although the WTD relations only hold for unidirectional transport. When the renewal assumption is violated, however, temporal correlations, quantifiable via the Pearson correlation coefficient, are present. {{} When the phonons are in equilibrium the Pearson correlation coefficient is formally zero at all voltages, hence subsequent waiting and first-passage times are uncorrelated. In the unequilibrated case, however, significant positive correlations exist for a strong electron-phonon coupling. As has been discussed in previous literature these positive correlations occur only in a small voltage range when an elastic shortcut channel opens. Due to the presence of bidirectional transitions, comparing the Fano factor and FPTD randomness parameter proved more accurate than the WTD randomness parameter at predicting non-renewal behaviour. In the last ten years, a comprehensive framework of fluctuation statistics for mesoscopic transport has been developed. Looking forward, we expect that these three, the FCS, WTD, and FPTD, will continue to play large role in analysing and describing single electron transport.}

\section{Acknowledgements}

This work was supported by an Australian Government Research Training Program Scholarship to S.L.R. 



%

\end{document}